# High-resolution luminescence spectroscopy of $CaWO_4:Ho^{3+}$: Sensitive detection of internal strains, temperature, and magnetic field

M. N. Popova[1,*], M. Diab[1,2], N. N. Kuzmin[1], K. A. Subbotin[3], A. I. Titov[3], B. Z. Malkin[4]

[1]*Institute of Spectroscopy, Russian Academy of Sciences, Troitsk, Moscow, 108840, Russia*
[2]*Moscow Institute of Physics and Technology (National Research University), Dolgoprudny, Moscow region, 141700, Russia*
[3]*Prokhorov General Physics Institute of the Russian Academy of Sciences, Moscow, 119991, Russia*
[4]*Kazan Federal University, Kazan, 420008, Russia*

## Abstract

We carried out high-resolution (0.02 $cm^{-1}$) measurements of the photoluminescence (PL) spectra of $CaWO_4:Ho^{3+}$, including in magnetic field, and performed spectra modeling. Well-resolved hyperfine structure was observed in the luminescence spectra of $CaWO_4:Ho^{3+}$ for the first time. Energies of several crystal-field levels were specified, *g*-factors were determined. Variations of a magnetic field as small as several tenths of mT can be detected from the PL spectra. We show that the PL spectra are much more sensitive to lattice strains in $CaWO_4:Ho^{3+}$ than the absorption spectra. The spectra were simulated using refined crystal-field parameters and taking into account both hyperfine and deformation interactions. Here, we used the previously obtained distribution function of random strains induced by point lattice defects in an elastically anisotropic $CaWO_4$ crystal. A luminescence thermometer for temperatures around 20 K based on $CaWO_4:Ho^{3+}$ is suggested.

* Corresponding author
E-mail: popova@isan.troitsk.ru
Phone: +7 495 851-02-34
Postal address:
Prof. Marina Popova
Institute of Spectroscopy RAS,
5, Fizicheskaya Str., Troitsk, Moscow, 108840, Russia

## I. INTRODUCTION

Calcium tungstate $CaWO_4$ has the scheelite-type tetragonal crystal structure (space group $I4_1/a$) and is characterized by chemical, mechanical and thermal stability and excellent optical properties. Both in pure form and doped with rare-earth (RE) ions, it is an important functional material with a wide range of applications. It is used in various fields such as lasers [1-3], scintillators [4,5], phosphors [6-9], including those transforming the UV-emission of GaN-based LEDs to a visible light [9-11], display devices [9,11], photocatalysis [12,13], anti-counterfeiting applications [14,15], medical devices, optoelectronics [16], luminescence thermometry [17-26], as well as photovoltaics [27,28]. The $CaWO_4$ matrix is characterized by a low density of nuclear spins, which leads to long coherence times of the hyperfine sublevels of RE ions embedded into the crystal [29,30]. In addition, in the case of $CaWO_4$, it is possible to obtain a crystal free of nuclear spins if isotopically enriched raw materials are used. These circumstances have caused great interest in calcium tungstate as a material for modern quantum technologies [29-37]. The knowledge of hyperfine structure (HFS) of crystal-field (CF) levels for RE ions doped into $CaWO_4$ is essential in this case. HFS in luminescence spectra can also be used to implement sensitive sensors of magnetic field [38], temperature [39], and strain [40], as has been shown by the example of $^7LiYF_4:Ho^{3+}$ [38-40]. In that case, crystals isotopically pure in lithium were used to avoid additional spectral structure caused by isotopic disorder ($^7Li$ - $^6Li$) in the lithium sublattice [41]. Unlike $LiYF_4$, a heterogeneous isotopic composition in $CaWO_4$ is not important because the relative mass defects in the calcium and tungsten sublattices are small as compared to the relative mass differences of $^7Li$ and $^6Li$ in $LiYF_4$.

Holmium is well-known laser ion used primary for generation of the emission above 2 μm ($^5I_7 \rightarrow {}^5I_8$ transition) promising for the applications in medicine [42], remote gas sensing and wind mapping [43,44], as well as for pumping of mid-infrared lasers and parametric oscillators [45]. During the last years an efficient lasing in various regimes at $Ho^{3+}$ doped crystals and fibers was reported [46-49] including the lasing at $Ho^{3+}$ doped Scheelite-like tungstate and molybdate crystals [50-52]. Besides that, various up- and downconversion schemes (including those aimed for temperature sensing) have been realized in phosphors based on $CaWO_4$ [18] and other Scheelite-like crystals [53-55] co-doped by $Ho^{3+}$ and some other RE ions. The $Ho^{3+}$ ion is characterized by well-resolved crystal-field structure in its optical absorption and emission spectra, even in partially disordered crystals like $CaWO_4$, and the largest among RE ions magnetic hyperfine constant, which is favorable for the observation of HFS in the spectra.

In earlier paper [56], well-resolved HFS was observed in optical absorption and EPR spectra of $CaWO_4:Ho^{3+}$ and the influence of random lattice strains on the HFS was investigated. In this work, we have undertaken high-resolution photoluminescence (PL) study of $CaWO_4:Ho^{3+}$, with the aim to explore possibilities of creating luminescent temperature, strain, and magnetic field sensors. Well-resolved hyperfine structure in the PL spectra of $CaWO_4:Ho^{3+}$ was observed for the first time. Behavior of the spectra in an external magnetic field was studied. Crystal-field calculations were performed and the shape of spectral lines was simulated, taking into account hyperfine interactions and random lattice strains. We show that the high-resolution luminescence spectra can serve as a sensitive tool for detecting internal deformations of the crystal lattice. Magnetic field and temperature luminescence sensors can be implemented on the basis of $CaWO_4:Ho^{3+}$.

## II. EXPERIMENTAL DETAILS

For $CaWO_4:Ho^{3+}$ (0.01 and 0.5 at. %) crystal growth, charges were prepared by mixing the initial chemicals, $CaCO_3$ (purity: 4 N, Unichim, USSR), $WO_3$ (5 N, Anhui Toplus Impex Co., LTD, China), $Ho_2O_3$ (4 N, USSR), and $Nb_2O_5$ (4 N, Krasnyi Khimik, USSR). $Nb^{5+}$ ions were added to the crystal composition as the charge compensators of heterovalent substitution of $Ca^{2+}$ by $Ho^{3+}$. $Nb^{5+}$ and $Ho^{3+}$ were taken in the equimolar amounts, 0.01 at. % and 0.5 at. % (for crystals with the corresponding holmium content) relative to the $Ca^{2+}$ content. The initial chemicals were weighted in the proper amounts at an Adventurer AX523 electronic analytical balance (OHAUS, USA) with the precision of ±0.01 g. Before the weighting the initial chemicals were pre-dried by calcining at 680 °C during 5 h. The thorough mixing was performed at a Multi RS-60 mixer-rotator (BioSan, Latvia) during 5 h. The mixture was then heated at an EKPS-10/1250 SPU 4107 muffle furnace (Russia) at 700 °C during 5 h in order to perform solid-phase synthesis of the target compound.

The crystals were grown from the melt by the Czochralski method at a Kristall-2 growth machine (USSR) from a Pt/Rh crucible with the sizes of ∅30×30 mm in air atmosphere. The single-crystalline bar of undoped $CaWO_4$ cut along [100] crystallographic axis was used as the seed. The pulling and rotation rates were 1 mm/h and 6 rpm, respectively. After finishing the crystal growth process and removal of the grown crystal from the melt surface it was slowly (at the rate of 8 °C/h) cooled to room temperature in order to minimize the cracking probability. The obtained crystals were transparent and almost colorless.

The refinement of the orientation of the grown crystals in respect to its $S_4$ optical axis with the precision of ±0.5° was performed by observation the crystal through the crossed polarizers at Biomed-5 optical polarizing microscope (LOMO, Russia). Two pairs of polished flat facets have been made at the grown crystals: one pair of the facets contained the optical axis of the crystal; another pair was perpendicular to this axis. The actual $Ho^{3+}$ concentrations in the crystals have not been measured. However, taking into account the literature data [28,57], as well as the fact of the usage of $Nb^{5+}$ as the charge compensator, one can evaluate the actual $Ho^{3+}$ content in the crystals as close to the nominal ones, i.e., ~ 0.01 at. % and 0.5 at. %.

High-resolution (up to 0.02 $cm^{-1}$) photoluminescence (PL) spectra were measured on an experimental setup built on the basis of a Bruker IFS 125 HR high-resolution Fourier spectrometer. PL was excited by a modified diode laser with wavelength 641.6 nm. A filter before the sample reduced the thermal load on the sample. An InSb and a high-gain InGaAs detectors to record the spectra and $CaF_2$ as a beam splitter for the interferometer were used. The $CaWO_4:Ho^{3+}$ sample was placed into a Sumitomo RP-082 closed-cycle helium cryostat with an electromagnet and a concentrating magnetic system attached directly to the first stage of the cryostat. A computer-controlled multi-channel current source Korad KA3305P was used to change the magnitude of the applied magnetic field. The magnetic field was directed along the *c* ($S_4$) axis of the crystal and could be varied from 0 to 500 mT. The temperature could be varied in the range 4 – 80 K, it was controlled and recorded using a LakeShore Model-335 PID temperature controller with an accuracy of ±0.05 K. The setup as well as the diode laser used for PL excitation are described in detail in our recent paper [38].

## III. EXPERIMENTAL RESULTS

### A. Photoluminescence of $CaWO_4:Ho^{3+}$ in zero magnetic field: crystal-field levels and hyperfine structure

Figure 1a shows the low-temperature PL spectrum of $CaWO_4:Ho^{3+}$ (0.01 at. %) in the entire spectral range studied. The scheme of energy levels of $Ho^{3+}$given in the Inset clarifies the identification of the observed optical manifolds. The spectra are not corrected for detector sensitivity, which increases smoothly by a factor of about 1.8 in the spectral interval from 0.95 to 1.67 μm and then drops. All observed lines can be identified as associated with $Ho^{3+}$ ions at the $S_4$ tetragonal symmetry sites with non-local charge compensation when replacing $Ca^{2+}$ with $Ho^{3+}$. Figure 1b, representing as an example the PL spectral manifold $^5I_7 \rightarrow {}^5I_8$ in expanded scale, illustrates this statement. The designations $n_i \rightarrow n_f$ in Fig. 1b indicate the initial and final levels of the corresponding transition, see the scheme of the Inset.

The CF energy levels of $Ho^{3+}$ in tetragonal sites with non-local charge compensation are characterized by the one-dimensional $\Gamma_1$ and $\Gamma_2$ and two-dimensional $\Gamma_{34}$ irreducible representations (IR) of the $S_4$ point symmetry group. Table 1 lists the energies and symmetries (IRs) of several CF levels of $Ho^{3+}$ in tetragonal sites of $CaWO_4$ (relevant for further discussion), determined from the analysis of polarized absorption [56] and luminescence spectra, using the selection rules (see the Supplemental Material [58], Tables S1 and S2). In the Supplemental Material [58], an extended Table S3 of CF levels and a Table S4 clarifying the lines' identification in Fig.1b, as well as the spectral manifolds $^5F_5 \rightarrow {}^5I_6$ and $^5I_6 \rightarrow {}^5I_8$ in expanded scale (Figures S2 and S3, respectively) and corresponding Tables S5 and S6 for the lines' identification are presented. In this study, we have specified several CF energies and added the missing $\Gamma_1^1$ level of the $^5I_6$ CF manifold.

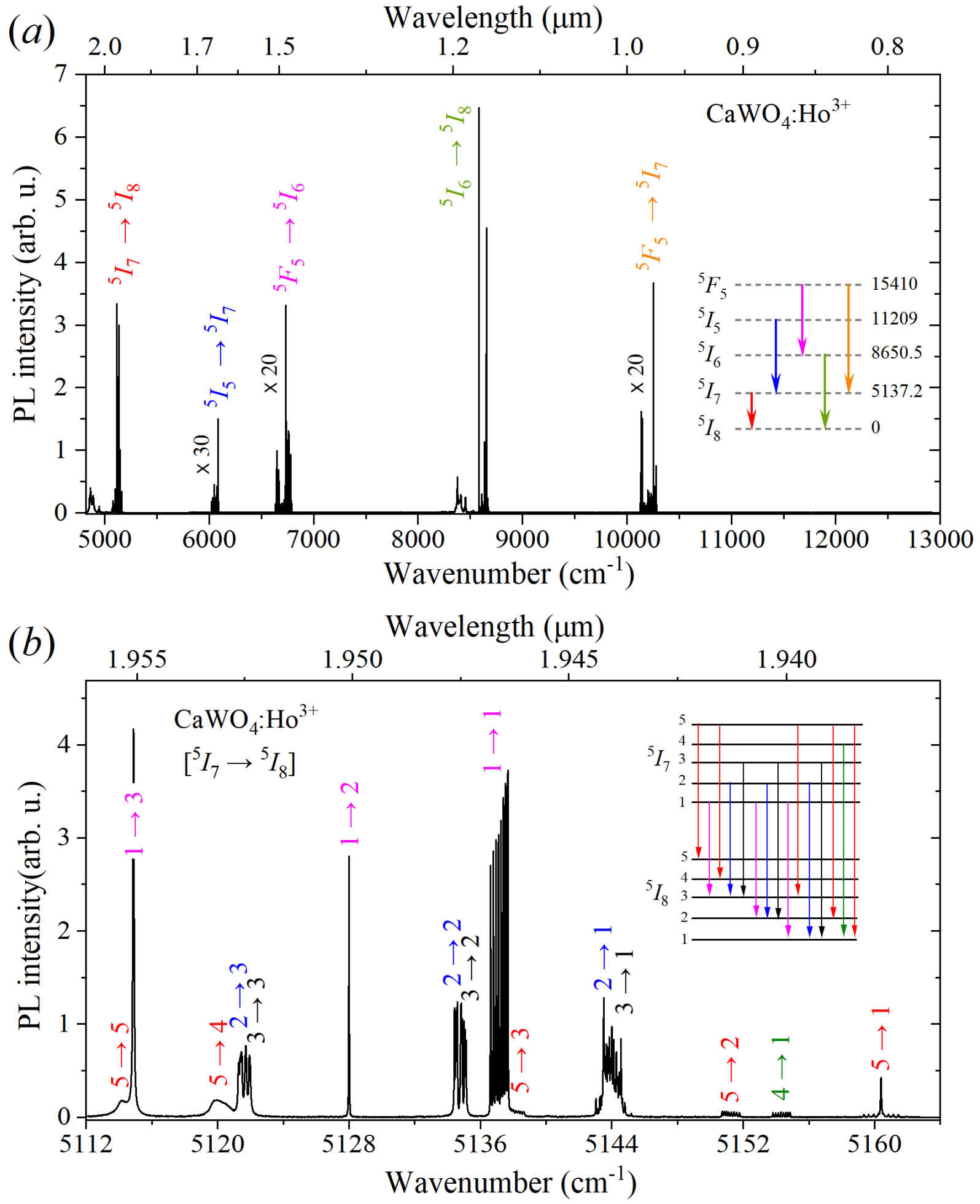

FIG. 1. PL spectrum of $CaWO_4$:$Ho^{3+}$ (0.01 at. %): $T = 15$ K; $\lambda_{ex} = 641.6$ nm. (a) The entire spectral range studied. Identified intermultiplet transitions of $Ho^{3+}$ are indicated. Inset presents the scheme of energy levels of $Ho^{3+}$; the observed transitions are shown by arrows. (b) The $^5I_7 \rightarrow {}^5I_8$ luminescent transition on an enlarged scale with HFS clearly observed. The numbers $n_i \rightarrow n_f$ indicate the initial and final levels of the corresponding transition, numbered from the lowest level in the CF multiplet (see Inset, Table 1 and Table S4 in the Supplemental Material [58]).

Table 1. Measured ($E_{exp}$) and calculated ($E_{theor}$) energies of the CF levels of $Ho^{3+}$ in $CaWO_4:Ho^{3+}$ and total hyperfine splittings $\Delta_{HF}$ and $g$-factors of $\Gamma_{34}$ non-Kramers doublets ([56] and present work)

| Level | | IR | $E_{exp}$ (cm$^{-1}$) | $E_{theor}$ (cm$^{-1}$) | $\Delta_{HF}$ (cm$^{-1}$) exp | $\Delta_{HF}$ (cm$^{-1}$) theor | $g_{\parallel}$ exp | $g_{\parallel}$ theor |
|---|---|---|---|---|---|---|---|---|
| $^5F_5$ | 8 | $\Gamma_{34}$ | 15585 | 15590 | - | 0.67 | - | 11.29 |
| | 7 | $\Gamma_1$ | 15568 | 15566 | | | | |
| | 2 | $\Gamma_{34}^1$ | 15427.5 | 15425 | - | 0.26 | - | 4.56 |
| | 1 | $\Gamma_2^1$ | 15410.1 | 15421 | | | | |
| | | | | | | | | |
| $^5I_6$ | 7 | $\Gamma_2^3$ | 8680.5 | 8676.2 | | | | |
| | 6 | $\Gamma_1^2$ | 8675.1 | 8670.4 | | | | |
| | 5 | $\Gamma_{34}^2$ | 8669.8 | 8662.9 | 0.37 | 0.32 | 3.4 | 3.03 |
| | 4 | $\Gamma_{34}^1$ | 8658.2 | 8655.1 | 1.07 | 1.03 | 9.9 | 9.93 |
| | 3 | $\Gamma_2^2$ | 8657.4 | 8660.5 | | | | |
| | 2 | $\Gamma_1^1$ | 8650.6 | 8650.7 | | | | |
| | 1 | $\Gamma_2^1$ | 8650.5 | 8650.5 | | | | |
| | | | | | | | | |
| $^5I_7$ | 5 | $\Gamma_{34}^2$ | 5160.5 | 5166.5 | 1.05 | 1.04 | 11.1 | 11.86 |
| | 4 | $\Gamma_1^1$ | 5154.4 | 5153.5 | | | | |
| | 3 | $\Gamma_2^2$ | 5144.0 | 5145.5 | | | | |
| | 2 | $\Gamma_{34}^1$ | 5143.9 | 5143.7 | 0.69 | 0.68 | 6.4 | 7.76 |
| | 1 | $\Gamma_2^1$ | 5137.2 | 5138.1 | | | | |
| | | | | | | | | |
| $^5I_8$ | 6 | $\Gamma_{34}^2$ | 67.8 | 71.0 | 0.70 | 0.75 | - | 9.73 |
| | 5 | $\Gamma_1^2$ | 46.2 | 53.1 | | | | |
| | 4 | $\Gamma_1^1$ | 40.5 | 46.7 | | | | |
| | 3 | $\Gamma_2^2$ | 22.2 | 20.8 | | | | |
| | 2 | $\Gamma_2^1$ | 9.2 | 9.2 | | | | |
| | 1 | $\Gamma_{34}^1$ | 0 | 0 | 1.05 | 1.05 | 13.9 | 13.69 |

In Fig. 1b, spectral lines with well-resolved HFS are clearly seen. The resolved HFS in optical *absorption* [56] and EPR [56,59] spectra of $CaWO_4:Ho^{3+}$ was observed and studied earlier. Here, we report on the first observation of HFS in the *luminescence* spectra of $CaWO_4:Ho^{3+}$. In a zero magnetic field, the $\Gamma_{34}$ CF levels possess an eight-component HFS resulting from the interaction of 4f electrons with the magnetic dipole and electric quadpupole moments of the holmium nucleus with spin $I = 7/2$. Each hyperfine component is doubly degenerate in zero magnetic field, the states $|\Gamma_3, m\rangle$ and $|\Gamma_4, -m\rangle$ have the same energy (here, $m$ is the component of nuclear moment $\boldsymbol{I}$ along the crystallographic $c$ axis, $-7/2 \le m \le 7/2$ ). For the $\Gamma_1$ and $\Gamma_2$ non-degenerate electronic CF states magnetic HFS is forbidden in the first approximation. Electric quadrupole and pseudoquadrupole (magnetic dipole in the second approximation) hyperfine interactions split $\Gamma_1$ and $\Gamma_2$ singlets into four nonequidistant hyperfine sublevels and lead to nonequidistance in $\Gamma_{34}$ hyperfine manifolds [60,61].

Figure 2 shows several lines with resolved HFS in the low-temperature PL spectrum of $CaWO_4:Ho^{3+}$. In the $^5I_7 \rightarrow {}^5I_8$ transition, which is allowed as a magnetic dipole one in the free $Ho^{3+}$ ion, magnetic dipole and electric dipole transitions have comparable intensities (pay attention to a strong line 1→3 in Fig. 1b corresponding to a purely magnetic dipole $\Gamma_2 \rightarrow \Gamma_2$ transition). As a result, HFS of the lines corresponding to $\Gamma_{34} \rightarrow \Gamma_{34}$ transitions consists of a superposition of the sum and difference of HFSs of the levels participating in the transition (Fig. 2a). As follows from simulations (see Section IV), the central intense line in Fig. 2a corresponds to electric dipole transitions; it has unresolved HFS, the intervals of which are the difference of almost equal intervals in HFSs of the levels $\Gamma_{34}^2(^5I_7)$ and $\Gamma_{34}^1(^5I_8)$. Another feature in the $^5I_7 \rightarrow {}^5I_8$ transition is connected with the fact that the CF levels 2 ($\Gamma_{34}$) and 3 ($\Gamma_2$) of $^5I_7$ have almost the same energy, which intensifies their interaction. This leads to a peculiar structure of the line in Fig. 2b. Figures 2c, 2d, and especially 2e, all of them for the $\Gamma_2 \rightarrow \Gamma_{34}$ transitions, demonstrate a regular nonequidistance of HFS typical for the influence of quadrupole and pseudoquadrupole hyperfine interactions. The final level for these lines is the ground level $\Gamma_{34}$. Different shapes of these lines are due to different contribution of quadrupole and pseudoquadrupole hyperfine interactions into HFS of the initial singlet CF levels of corresponding transitions. The same as in the absorption spectra of $CaWO_4:Ho^{3+}$ crystals with holmium concentration less than 0.5 at. % [56], no signatures of random lattice deformations are observed. In contrast, the line of the $^5F_5$ $\Gamma_2$ (15410.1 cm$^{-1}$) → $^5I_6$ $\Gamma_{34}$ (8669.8 cm$^{-1}$) transition between the excited multiplets, which is absent in the absorption spectra, demonstrates an enlarged central interval of HFS (see Fig. 2f). This is the signature of random lattice deformations [56,62,63]. Thus, PL spectra are more sensitive to random lattice strains than absorption spectra and can be used to implement a sensitive sensor of lattice deformations.

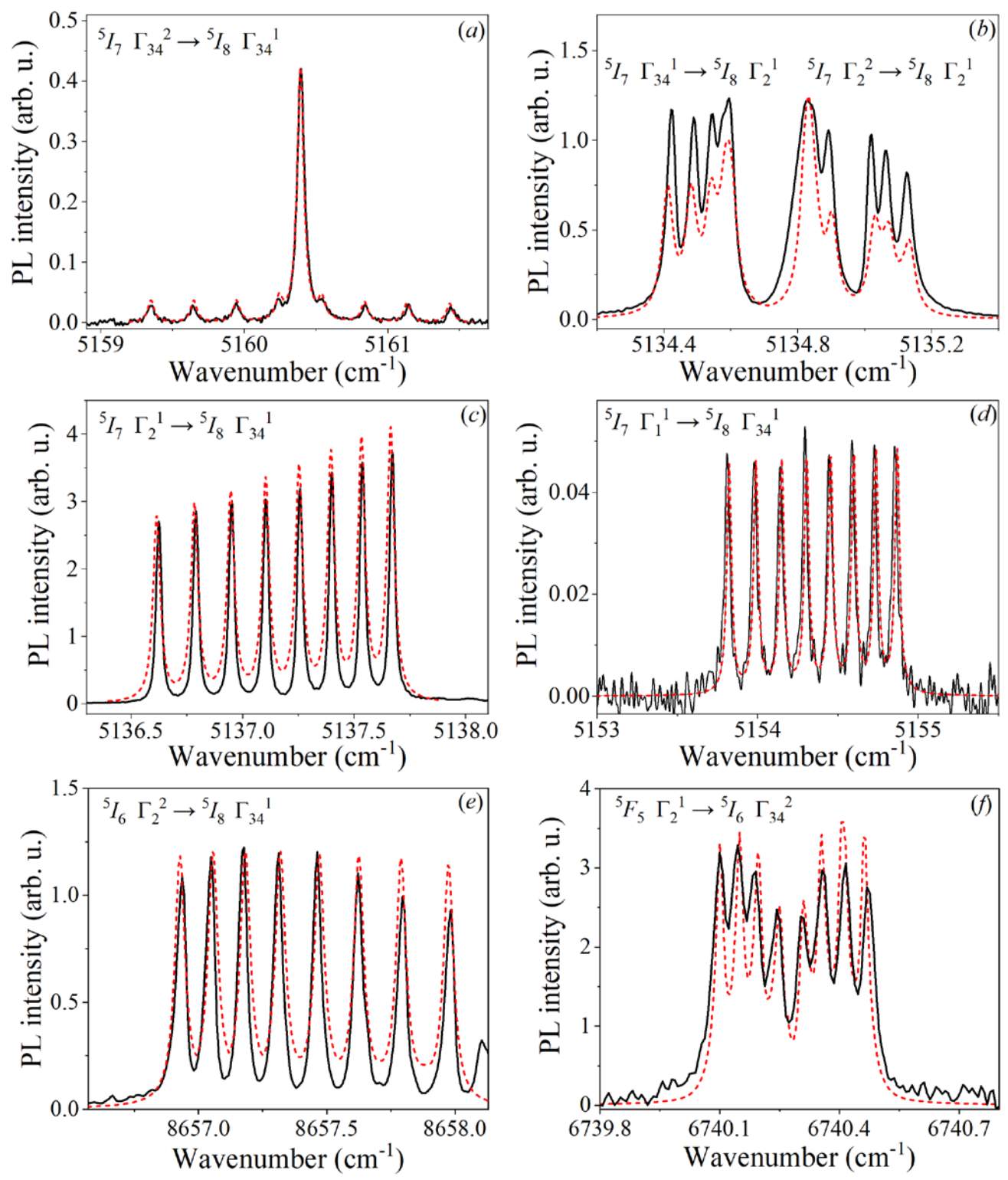

FIG. 2. Measured (solid black lines) and simulated (dash red lines) envelops of spectral lines with HFS in the PL spectrum of $CaWO_4:Ho^{3+}$(0.01%) in a zero magnetic field. $T$=15K; $\lambda_{ex}$ = 641.6 nm.

### B. Luminescence cryothermometry with $CaWO_4:Ho^{3+}$

The temperature-dependent spectra of $CaWO_4:Ho^{3+}$ indicate the possibility of realizing a luminescent thermometer for cryogenic temperatures. A recent paper [65] communicated a luminescent cryothermometer based on $Ho^{3+}$ ions in $KY_3F_{10}:Ho^{3+}$. Here, we have chosen two isolated lines, 8640.3 cm$^{-1}$ and 8582.7 cm$^{-1}$, corresponding to the transitions $^5I_6\ \Gamma_2^3$ (8680.5 cm$^{-1}$) → $^5I_8\ \Gamma_1^1$ (40.5 cm$^{-1}$) and $^5I_6\ \Gamma_2^1$ (8650.5 cm$^{-1}$) → $^5I_8\ \Gamma_{34}^2$ (67.8 cm$^{-1}$), respectively, in the luminescence spectrum of $CaWO_4:Ho^{3+}$ to study the temperature dependence of their relative intensities. The initial levels of these transitions are separated by the energy interval $\Delta E = E_2 - E_1 = 30$ cm$^{-1}$, and if the Boltzmann distribution of their populations is realized, i.e. $\frac{n_2(T)}{n_1(T)} = \frac{g_2}{g_1} e^{-\Delta E/kT}$, the line intensity ratio ($LIR$) obeys the following relation:

$$LIR(T) = \frac{I_2(T)}{I_1(T)} = \frac{W_2\, n_2(T)}{W_1\, n_1(T)} = C\, e^{-\Delta E/kT}, \tag{1}$$

where $C = \frac{g_2}{g_1}\frac{W_2}{W_1}$, $g_2$ and $g_1$ are the degeneracy factors of the levels 2 and 1, $W_2$ and $W_1$ are radiative transition probabilities. In order to recommend the pair of luminescence lines for the Boltzmann ratiometric thermometry, it is necessary to check whether Eq. (1) is fulfilled. Figure 3 shows the two considered luminescence lines at several temperatures and the temperature dependence of their relative intensities compared with the Boltzmann distribution for the population of initial levels of the corresponding optical transitions. It can be seen that the ratio of the lines intensities is proportional to the ratio of Boltzmann populations and, thus, the luminescence Boltzmann ratiometric thermometer can be realized.

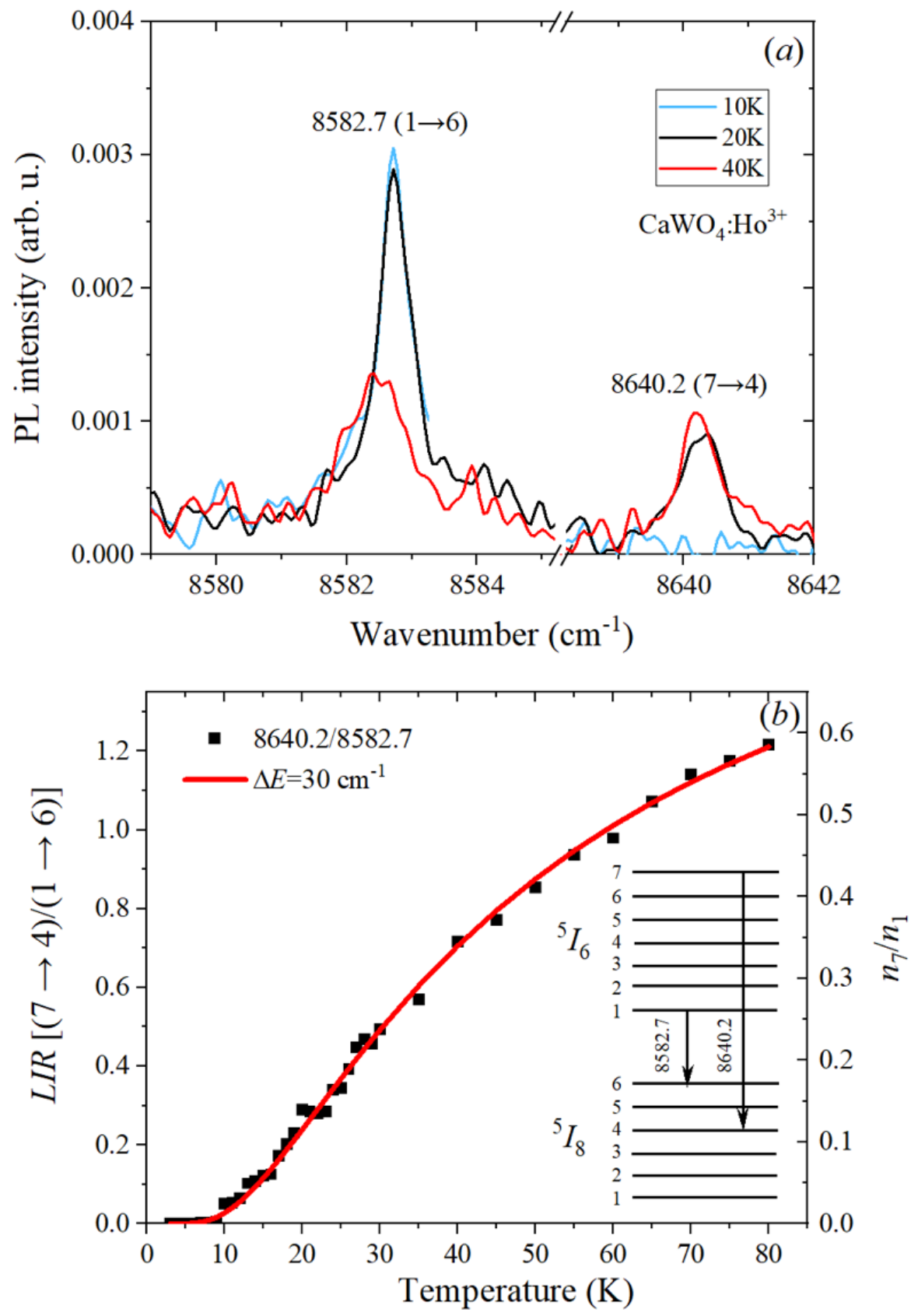

FIG. 3. (a) PL spectra of $CaWO_4$:$Ho^{3+}$(0.5%) at different temperatures in the region of the transitions 1 → 6 [$^5I_6$ $\Gamma_2^1$ (8650.5 cm$^{-1}$) → $^5I_8$ $\Gamma_{34}^2$ (67.8 cm$^{-1}$)] and 7 → 4 [$^5I_6$ $\Gamma_2^3$ (8680.5 cm$^{-1}$) → $^5I_8$ $\Gamma_1^1$ (40.5 cm$^{-1}$)]; $\lambda_{ex}$ = 641.6 nm; spectral resolution is 0.2 cm$^{-1}$. (b) Temperature dependences of *LIR* [(7 → 4)/(1 → 6)] of the integrated intensities of the luminescence lines at 8640 cm$^{-1}$ and 8682.7 cm$^{-1}$ (black symbols), and the population ratios $n_7/n_1$ of the levels 7 and 1, separated by an interval of 30 cm$^{-1}$, assuming a Boltzmann distribution (red line).

In this case, the absolute thermal sensitivity $S_a(T)$

$$S_a(T) = \frac{d\,LIR(T)}{d\,T} = C\,\frac{d(e^{-\Delta E/kT})}{d\,T} = C\,\frac{\Delta E}{kT^2}e^{-\Delta E/kT} \tag{2}$$

has a maximum at temperature $T_m = \Delta E/2k$.

The relative thermal sensitivity $S_r(T)$ (used to compare thermometers based on different principles) is:

$$S_r(T) = \frac{1}{LIR(T)} \frac{d\, LIR(T)}{d\, T} = \frac{S_a(T)}{LIR(T)} = \frac{\Delta E}{kT^2} \quad (3)$$

Figure 4 presents the temperature dependences of these quantities for the case of selected lines in the luminescence spectra of $CaWO_4{:}Ho^{3+}$. The best absolute thermal sensitivity is achieved at the temperature $T_m$ = 21.5 K, the relative sensitivity at this temperature is 10 % $K^{-1}$.

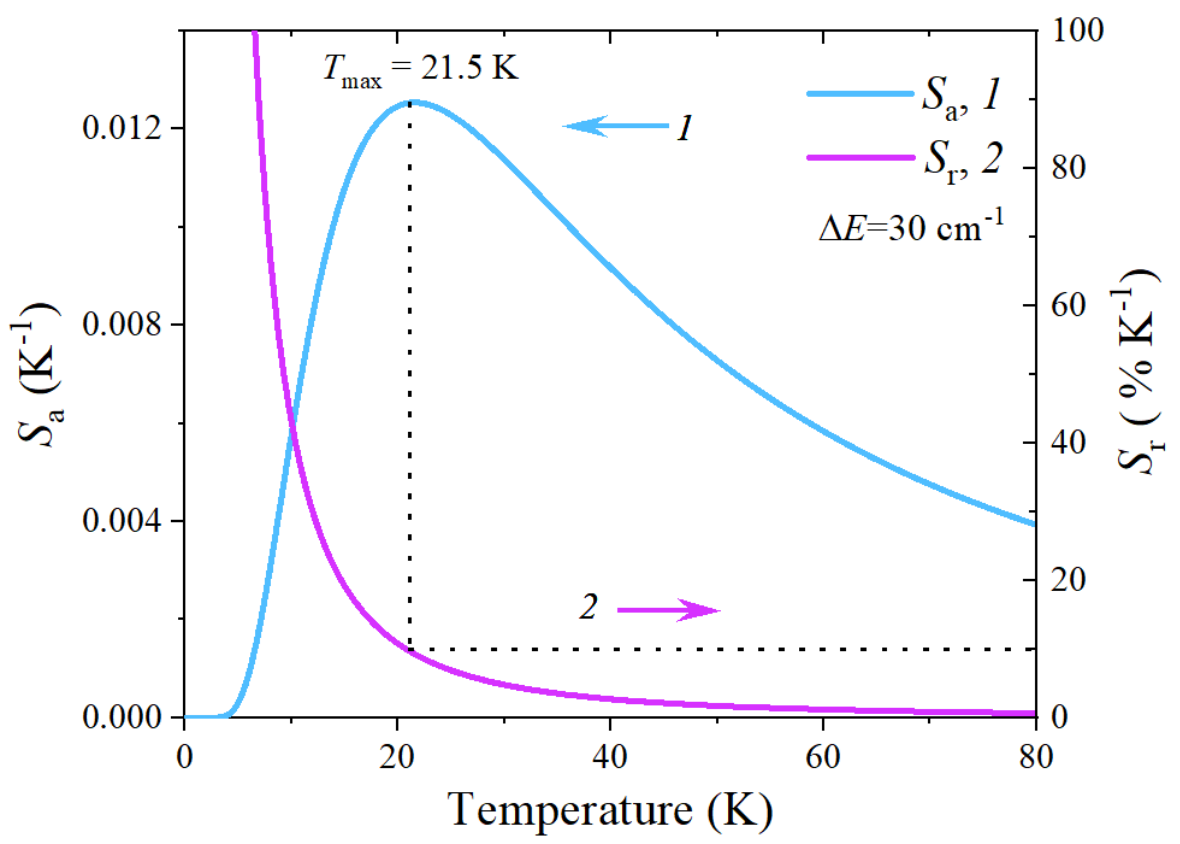

FIG. 4. The absolute $S_a$ (blue curve) and relative $S_r$ (magenta curve) sensitivities for a luminescence Boltzmann ratiometric thermometer with $\Delta E$ = 30 cm$^{-1}$.

We also checked the possibility to use the PL lines, the initial levels of which are separated by a smaller interval $\Delta E$. The best achieved result is illustrated at Fig. 5 where line intensity ratio $LIR(T)$ for the PL lines with $\Delta E$ = 7.7 cm$^{-1}$ is compared with the ratio of Boltzmann populations. Those are the PL lines 8658.2 cm$^{-1}$ and 8650.5 cm$^{-1}$ corresponding to the transitions $^5I_6\ \Gamma_{34}^2$ (8669.8 cm$^{-1}$) → $^5I_8\ \Gamma_{34}^1$ (0 cm$^{-1}$) and $^5I_6\ \Gamma_2^1$ (8650.5 cm$^{-1}$) → $^5I_8\ \Gamma_{34}^1$ (0 cm$^{-1}$), respectively. A systematic deviation of experimental points from the theoretical curve is observed at temperatures below 10 K (from 0.2 K at 10 K to 0.5 K at 3 K). A significantly greater deviation was observed without filters that weaken the laser excitation intensity. Thus, to use $CaWO_4{:}Ho^{3+}$ as a luminescent thermometer for temperatures below 10 K, it is necessary to exclude heating of the sample by laser radiation that excites luminescence. One of the possible ways is an excitation to a level close in energy to the initial levels of the luminescent transitions.

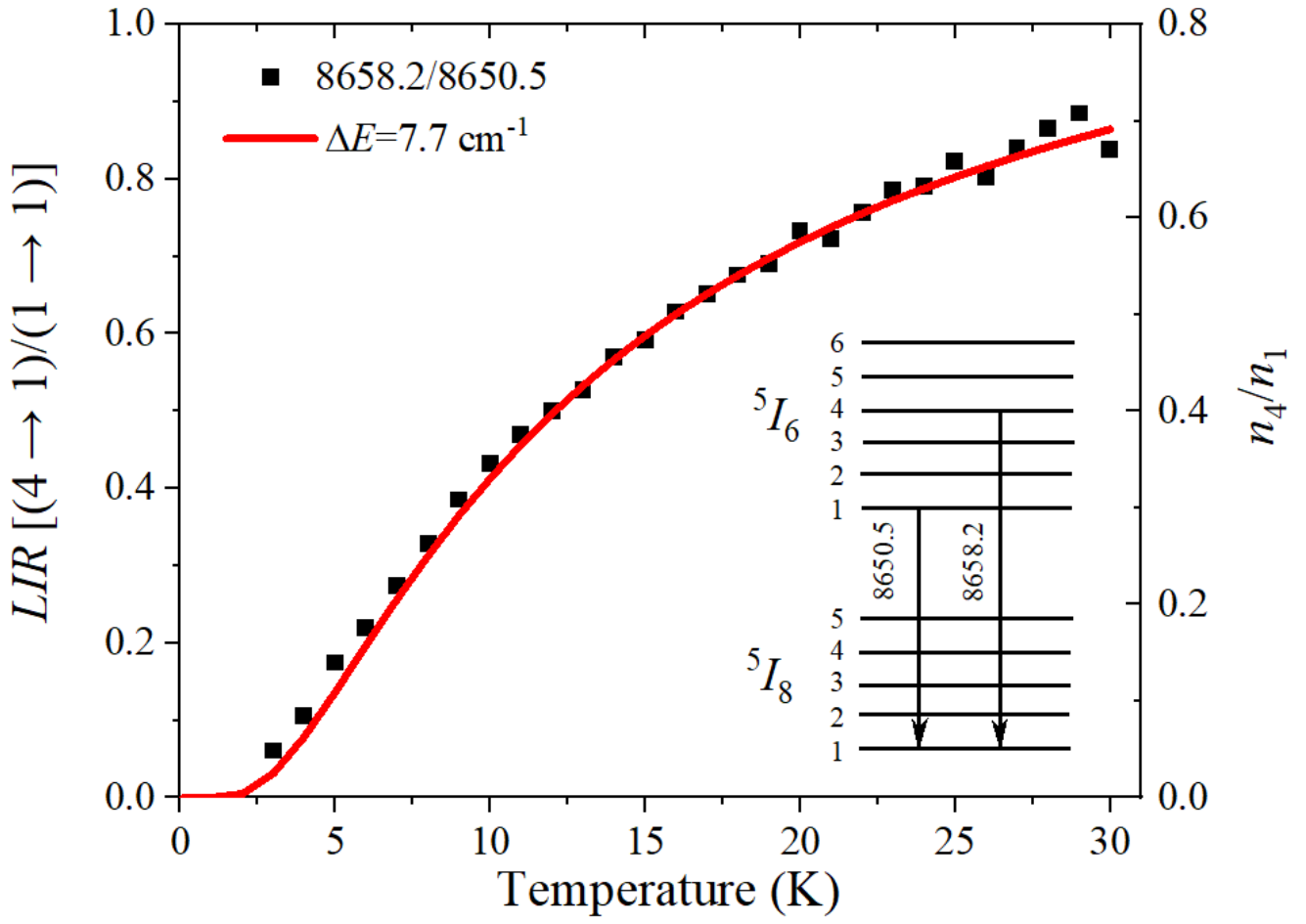

FIG. 5. Temperature dependences of *LIR* [(4 → 1)/(1 → 1)] of the integrated intensities of the luminescence lines at 8658.2 cm$^{-1}$ and 8650.0 cm$^{-1}$ (black symbols), and the population ratios $n_4/n_1$ of the levels 4 and 1, separated by an interval of 7.7 cm$^{-1}$, assuming a Boltzmann distribution (red line).

### C. Photoluminescence of $CaWO_4:Ho^{3+}$ in a magnetic field B||*c*

Color intensity map of Fig. 6a shows the magnetic-field behavior of the spectral line corresponding to the $^5I_7$ $\Gamma_2{}^1$ (5137.2 cm$^{-1}$) → $^5I_8$ $\Gamma_{34}{}^1$ (0) transition, which reflects HFS of the ground $\Gamma_{34}{}^1$ state of $CaWO_4:Ho^{3+}$ in a magnetic field. An external magnetic field parallel to the *c* axis of the crystal, **B**||*c*, lifts twofold degeneracy of the hyperfine states |$\Gamma_3$, *m*> and |$\Gamma_4$, -*m*> and splits each hyperfine level into two Zeeman sublevels separated by the interval $g_{||}\,\mu_B\,B$, where $\mu_B$ is the Bohr magneton and $g_{||}$ is the *g* factor of the electronic doublet. The hyperfine pattern of Fig. 6a is asymmetric: intervals in the high-frequency branch are markedly smaller than those in the low-frequency branch. This is due to the interaction with nearby-lying CF levels. CF calculations confirm such interpretation (see Fig. 6b, Fig. S4 in the Supplemental Material [58], and Section IV).

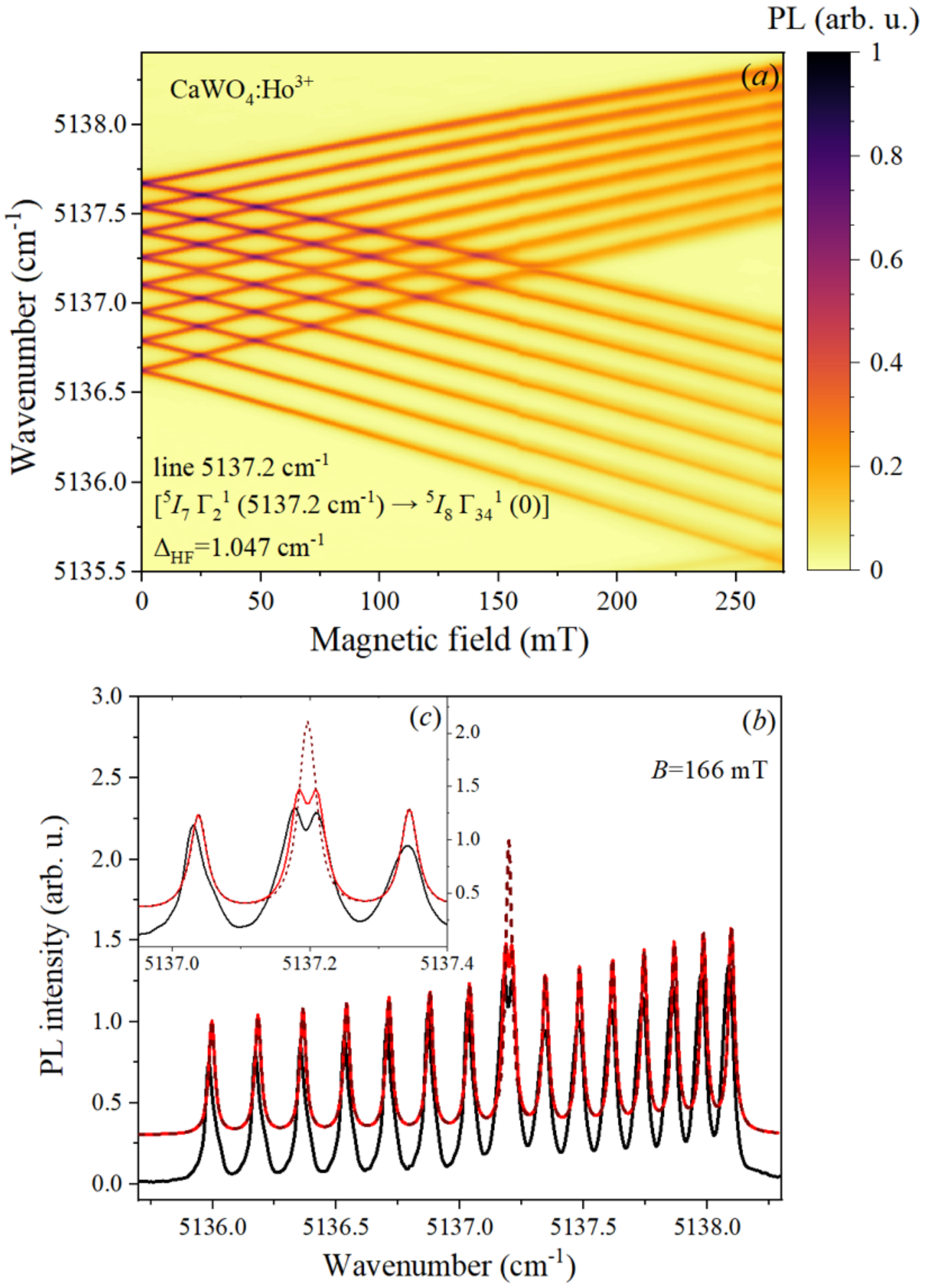

FIG.6. The $^5I_7\,\Gamma_2{}^1$ (5137.2 cm$^{-1}$) → $^5I_8\,\Gamma_{34}{}^1$ (0) luminescence transition of $CaWO_4$:$Ho^{3+}$ (0.01 at. %) in a magnetic field **B**||***c***. $T$ = 5 K; $\lambda_{ex}$ = 641.6 nm. (a) Luminescence intensity map in the magnetic-field – wavenumber scale; (b) Experimental (black) and calculated (red) spectra at $B$ = 166 mT. Dotted line is a spectrum calculated without taking into account electron-deformation interaction. Inset shows the region of the 4$^{th}$ anticrossing of the hyperfine levels with $\Delta m$=0 on an extended scale. A deformational gap of 0.025 cm$^{-1}$ is clearly observed.

Clearly visible gaps at the points of crossing of hyperfine components with $\Delta m$=0 should be noted. The intersecting hyperfine sublevels repel each other (anticrossing). Such gap at the fourth anticrossing point is shown in Fig. 6b and, at extended scale, in the Inset of Fig. 6b. Measured and calculated (see Section IV) spectra at the magnetic field values corresponding to all $\Delta m$=0 anticrossing points are presented in the Supplemental Material [58], Fig.S5. The $\Delta m$=0 anticrossings were observed in the EPR spectra of $CaWO_4$:$Ho^{3+}$(0.05%) [56] and in optical spectra of $^7LiYF_4$:$Ho^{3+}$ [64]. They were shown to arise due to random lattice strains always present in a real crystal [56,64]. We note that the optical absorption and emission spectra of $CaWO_4$ doped with holmium in amount of 0.01 at. % or even 0.05 at.% [56] do not show any signature of random lattice strains when registered at a zero magnetic field. Thus, Zeeman spectroscopy offers much more sensitive tool for detecting lattice deformations.

Hyperfine levels' anticrossings occur also for $|\Delta m| = 2$, i.e., gaps are observed at the points of crossing between the $|\Gamma_4, m>$ and $|\Gamma_3, m\pm 2>$ states. Two examples of rather large $|\Delta m| = 2$ gaps are presented in Fig. 7. These gaps are caused by the transverse term in magnetic dipole hyperfine interaction [64]. At the anticrossing point, the wave functions of the crossing levels are present with equal weights.

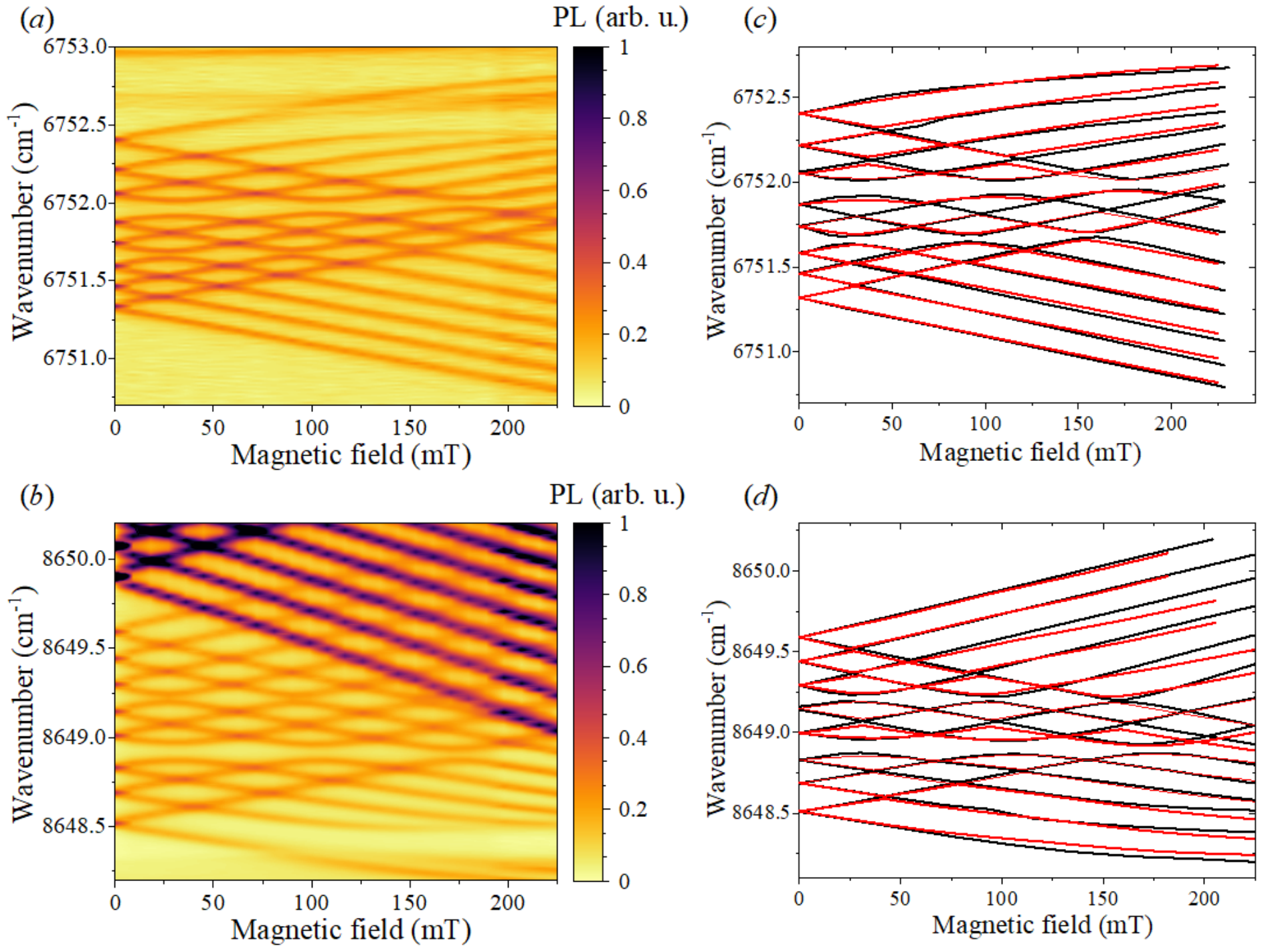

FIG. 7. (a,b) Luminescence intensity maps and (c,d) the corresponding evolution of the transition frequencies in the HFS (black and red lines show the results of measurements and simulations, respectively) of singlet-doublet transitions (a,c) $^5F_5\ \Gamma_2$ (15410.1 cm$^{-1}$) → $^5I_6\ \Gamma_{34}$ (8658.2 cm$^{-1}$) and (b,d) $^5I_6\ \Gamma_{34}$ (8658.2 cm$^{-1}$) → $^5I_8\ \Gamma_2$ (9.16 cm$^{-1}$) of $CaWO_4:Ho^{3+}$ (0.01 at. %) in a magnetic field **B**||***c***. $T$ = 15 K; $\lambda_{ex}$ = 641.6 nm. Large hyperfine ($\Delta m = \pm 2$) anticrossings are observed.

The measured dependences of the luminescence spectra on the magnetic field make it possible to create a sensitive luminescent sensor of changes in the magnetic field. To estimate the precision of such sensor, we consider the line at 5160.5 cm$^{-1}$ [$^5I_7\ \Gamma_{34}$ (5160.5) → $^5I_8\ \Gamma_{34}$ (0)], HFS of which consists of a superposition of the sum and difference of HFSs of the levels participating in the transition (see Fig. 2a, Section IIIA). Correspondingly, the $g$ factor for the summary HFS of the line equals to the sum of $g$ factors of the involved $\Gamma_{34}$ levels, $g_{||}$ = 13.9 + 11.1 = 25.0, whereas the $g$ factor for the difference HFS is a difference of the corresponding $g$ factors (2.8). Figure 8 shows the behavior of the 5160.5 cm$^{-1}$ PL line in a magnetic field ***B*||*c*.** The smallest change in magnetic field $\delta B$ that can be detected depends on the linewidth and for the case of Fig. 8 is estimated to be better than $\delta B$ = 0.8 mT (see the Supplemental Material [58] for more details).

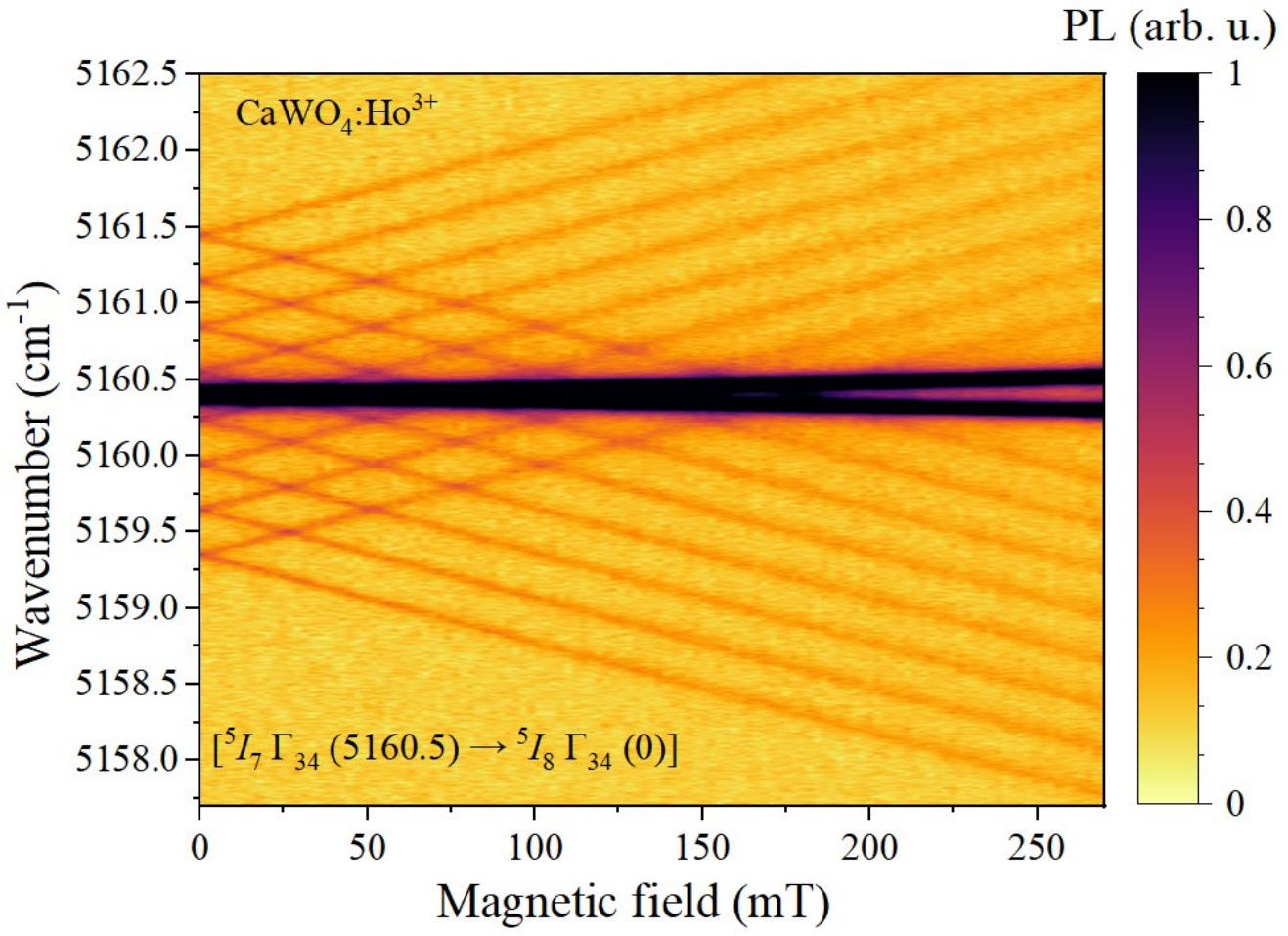

FIG. 8. Luminescence intensity map in the magnetic-field – wavenumber scale for the $^5I_7\,\Gamma_{34}$ (5160.5) → $^5I_8\,\Gamma_{34}$ (0) transition of $CaWO_4:Ho^{3+}$ (0.01 at. %) in a magnetic field **B**||***c***. $T$ = 15 K; $\lambda_{ex}$ = 641.6 nm. Intense central part corresponds to electric dipole transitions with unresolved difference HFS.

Figures S4 – S8 in the Supplemental Material [58] present several more examples of the HFS in the spectra of $CaWO_4:Ho^{3+}$ in an external magnetic field ***B***||$c$, with experimentally found $g$ factors.

## IV. SPECTRA MODELING

Modeling of the registered luminescence spectra of the $CaWO_4:Ho^{3+}$ crystal was based on the consideration of the $Ho^{3+}$ single-ion effective Hamiltonian

$$H = H_0 + H_{CF} + H_Z + H_{ed} + H_{HFM} + H_{HFQ} \quad (4)$$

where $H_0$ is the standard parameterized electronic Hamiltonian of a ‘free’ ion [66],

$$H_0 = \zeta \sum_k \boldsymbol{l}_k \boldsymbol{s}_k + \alpha \hat{\boldsymbol{L}}^2 + \beta \hat{G}(G_2) + \gamma \hat{G}(R_7) + \sum_q (F^q \hat{f}_q + P^q \hat{p}_q + T^q \hat{t}_q + M^q \hat{m}_q), \quad (5)$$

$H_{CF}$ is the energy of 4$f$ electrons in the crystal field of $S_4$ symmetry,

$$H_{CF} = \sum_k (B_2^0 O_{2,k}^0 + B_4^0 O_{4,k}^0 + B_6^0 O_{6,k}^0 + B_4^4 O_{4,k}^4 + B_4^{-4} O_{4,k}^{-4} + B_6^4 O_{6,k}^4 + B_6^{-4} O_{6,k}^{-4}), \quad (6)$$

$H_Z = -\boldsymbol{MB}$ is the Zeeman energy of a $Ho^{3+}$ ion in the external magnetic field $\boldsymbol{B}$, and $H_{ed}$ is the energy of interaction of 4*f* electrons with the field of crystal-lattice deformations in the linear approximation,

$$H_{\mathrm{ed}} = \sum_{k,pq} \sum_{\Gamma\lambda} B_p^q(\Gamma\lambda) O_{p,k}^q e(\Gamma\lambda), \tag{7}$$

and the operators $H_{HFM}$ and $H_{HFQ}$ correspond to magnetic dipole and electric quadrupole hyperfine interactions [67,68], respectively,

$$\begin{aligned} H_{\mathrm{HFM}} = A_{\mathrm{HF}} \sum_k \{ \boldsymbol{r}\boldsymbol{l}_k \boldsymbol{I} + \frac{1}{2}[O_{2,k}^0(3s_{kz}I_z - \boldsymbol{s}_k\boldsymbol{I}) + 3O_{2,k}^2(s_{kx}I_x - s_{ky}I_y) + 3O_{2,k}^{-2}(s_{kx}I_y + s_{ky}I_x) \\ + 6O_{2,k}^1(s_{kx}I_z + s_{kz}I_x) + 6O_{2,k}^{-1}(s_{kz}I_y + s_{ky}I_z)]\}, \end{aligned} \tag{8}$$

$$\begin{aligned} H_{\mathrm{HFQ}} = \frac{e^2Q(1-\gamma_\infty)}{4I(2I-1)} \sum_L q_L \frac{3Z_L^2 - R_L^2}{R_L^5} I_0 \\ - \frac{e^2Q(1-R_Q)}{4I(2I-1)} \left\langle \frac{1}{r^3} \right\rangle_{4f} \sum_k [O_{2,k}^0 I_0 + 3O_{2,k}^2 I_2 + 3O_{2,k}^{-2} I_{-2} + 6O_{2,k}^1 I_1 + 6O_{2.k}^{-1} I_{-1}]. \end{aligned} \tag{9}$$

All operators (5-9), as well as the magnetic moment operator of the $Ho^{3+}$ ion in an axial crystal field,

$$\boldsymbol{M} = -\mu_B \sum_k (\boldsymbol{r}\boldsymbol{l}_k + 2\boldsymbol{s}_k), \tag{10}$$

contain sums over ten 4*f* electrons with the orbital and spin moments $\boldsymbol{l}_k$ and $\boldsymbol{s}_k$, respectively, labeled by the index *k*. In (5-10), $\boldsymbol{L} = \sum_k \boldsymbol{l}_k$ and $\boldsymbol{S} = \sum_k \boldsymbol{s}_k$ are operators of the total single-ion orbital and spin moments, operators $O_{p,k}^q$ are linear combinations of the one-electron spherical tensor operators $C_p^{(q)}$ defined in [69] (note, that in the basis of the angular moment $\boldsymbol{J}$ eigenfunctions these operators coincide with the Stevens operators $O_p^q(\boldsymbol{J})$), $\boldsymbol{r}$ is the diagonal tensor of the orbital reduction factor [67], $r_{xx}=r_{yy}=r_\perp$, $r_{zz}=r_\parallel$ .

Values of parameters of the free-ion Hamiltonian (2) were presented earlier in our study of the optical absorption and EPR spectra of $CaWO_4:Ho^{3+}$ single crystals [56] (see also the Supplemental Material [58]). The initial values of CF parameters $B_p^q$ in the CF Hamiltonian (6) written in the Cartesian system of coordinates with *x, y* and *z* axes along the crystallographic axes (*a, b, c*) were calculated by making use of the semi-phenomenological exchange charge model (ECM) [70] (see Table 2). In these calculations, we used atomic coordinates from [71], charges $q_{Ca}$=2, $q_W$=3.8, and $q_O$=-1.45 (in units of the elemental charge) of point ions $Ca^{2+}$, $W^{6+}$, and $O^{2-}$, and model parameters $G_\sigma$=14, $G_\pi$=8 and $G_s$=12 which allowed us to reproduce the measured total CF splitting of the ground multiplet $^5I_8$.

Table 2. CF parameters $B_p^q$ (cm$^{-1}$) for $Ho^{3+}$ ions at the $Ca^{2+}$ sites in $CaWO_4$.

| *p* | *q* | [72]* | [56] | Present work** |
|---|---|---|---|---|
| 2 | 0 | 218 | 220 | 220 (234) |
| 4 | 0 | -83 | -80.5 | -79.5 (-98) |
| 4 | 4 | 814.7 (-586) | -593 | -593 (-573) |
| 4 | -4 | 0 (-566) | -582 | -582 (-515) |
| 6 | 0 | -2.06 | -2.345 | -2.545 (-5.85) |
| 6 | 4 | 391.5 (-377) | -304 | -314 (-289) |
| 6 | -4 | -137.5 ( -173) | -304 | -294 (-347) |

*Parameters in brackets were obtained by the rotation of the coordinate system around the *z*(*c*)-axis by the angle -56°.

**Parameters in brackets were calculated using the exchange-charge model in the crystallographic system of coordinates.

The space of six independent components of the symmetric deformation tensor $e_{\alpha\beta}$ of the scheelite lattice can be decomposed into subspaces transforming according to IR of the factor-group $C_{4h}$.

$$e\left(A_g^1\right)=e_1=\left(e_{xx}+e_{yy}+e_{zz}\right)\Big/\sqrt{6},\;\; e\left(A_g^2\right)=e_2=\left(2e_{zz}-e_{xx}-e_{yy}\right)\Big/2\sqrt{3}, \tag{11}$$

$$e\left(B_g^1\right)=e_3=\left(e_{xx}-e_{yy}\right)\Big/2,\quad e\left(B_g^2\right)=e_4=e_{xy},\quad e\left(E_g1\right)=e_5=e_{xz},\quad e\left(E_g2\right)=e_6=e_{yz}. \tag{12}$$

The distribution function of random strains induced by point lattice defects in an elastically anisotropic $CaWO_4$ crystal was obtained in [63] using the elastic constants $C_{11}$=145.9, $C_{33}$=127.4, $C_{12}$=62.6, $C_{13}$=39.2, $C_{16}$=-19.2, $C_{44}$=33.5, and $C_{66}$=38.7 (GPa), taken from [73]:

$$\begin{aligned} g(\boldsymbol{e}) = {} & \frac{15\nu_A\nu_B\xi}{8\pi^3\gamma_A^2\gamma_B^2\gamma_E^2}\times\Big\{\gamma_A^{-2}\left[\nu_A^2 e'\left(A_g^1\right)^2+e'\left(A_g^2\right)^2\right] \\ & +\gamma_B^{-2}\left[\nu_B^2 e'\left(B_g^1\right)^2+e'\left(B_g^2\right)^2\right]+\gamma_E^{-2}\sum_{\lambda=1,2}e\left(E_g\lambda\right)^2+\xi^2\Big\}^{-7/2}. \end{aligned} \tag{13}$$

Here,

$$e'(\Gamma^1)=\cos\psi_\Gamma e(\Gamma^1)+\sin\psi_\Gamma e(\Gamma^2),\;\; e'(\Gamma^2)=-\sin\psi_\Gamma e(\Gamma^1)+\cos\psi_\Gamma e(\Gamma^2),\;\; \Gamma^i = A_g \text{ or } B_g. \tag{14}$$

Rotations by angles $\psi_\Gamma$ in the planes $\{e(\Gamma^1), e(\Gamma^2)\}$ containing deformation tensor components which transform according to the same IR ($\Gamma^i =$ $A_g$ or $B_g$) are used to obtain the canonical form of the distribution function. The corresponding rotation angles equal $\psi_{A_g}$=-2.3° and $\psi_{B_g}$=-48.7°. The parameters $\nu_A$=5.56, $\nu_B$=2.44, $\gamma_A$=30.5, $\gamma_B$=43.8, $\gamma_E$=32.4 determine the effective widths of distributions $\gamma_\Gamma \xi / \nu_\Gamma$ for strains of different symmetry.

Totally symmetric random deformations $A_g^i$ shift CF sublevels of electronic multiplets and broaden the observed spectral lines, deformations of $E_g$ symmetry can be neglected within the first-order approximation, the fine deformational structure of non-Kramers doublets can only be induced by two rhombic deformations of $B_g$ symmetry.

The Hamiltonian of the electron-deformation interaction can be written in the form $H_{\mathrm{ed}} = \sum_{\alpha\beta} \hat{B}_{\alpha\beta} e_{\alpha\beta}$, where electronic operators $\hat{B}_{\alpha\beta} = \sum_{k,pq} B^q_{p,\alpha\beta} O^q_{p,k}$ are defined by the parameters [70]

$$B^q_{p,\alpha\beta} = \frac{1}{2}\sum_L (X_{L\alpha}\frac{\partial}{\partial X_{L\beta}} + X_{L\beta}\frac{\partial}{\partial X_{L\alpha}}) B^q_p(\boldsymbol{R}_L) \quad (15)$$

connected in Eq. (15) with the generalized CF parameters depending explicitly on radius-vectors $\boldsymbol{R}_L$ of the host lattice ions. Using the symmetrized linear combinations of deformations (11,12), we obtain explicit expressions for parameters to be introduced into Eq. (7):

$$B^q_p(A^1_g) = (2/3)^{1/2}(B^q_{p,xx} + B^q_{p,yy} + B^q_{p,zz}), \quad (16)$$

$$B^q_p(A^2_g) = (1/3)^{1/2}(2B^q_{p,zz} - B^q_{p,xx} - B^q_{p,yy}), \quad (17)$$

$$B^q_p(B^1_g) = B^q_{p,xx} - B^q_{p,yy}, \quad B^q_p(B^2_g) = 2B^q_{p,xy}. \quad (18)$$

Values of parameters (16-18) calculated in the framework of ECM [70] are presented in Table 3.

Table 3. Parameters of electron-deformation interaction (cm$^{-1}$).

| $p$ $q$ | $B^q_p(A_g{}^1)$ | $B^q_p(A_g{}^2)$ | $p$ $q$ | $B^q_p(B_g{}^1)$ | $B^q_p(B_g{}^2)$ |
|---|---|---|---|---|---|
| 2 0 | -618.2 | 1409.5 | 2 2 | 2182 | 3350 |
| 4 0 | 650 | 190 | 2 -2 | 3350 | -2423 |
| 4 4 | 4001 | 4196.3 | 4 2 | 966 | 2121 |
| 4 -4 | 3447 | 699 | 4 -2 | 2121 | -3123 |
| 6 0 | 41 | 99.5 | 6 2 | -282.5 | 248.2 |
| 6 4 | 2150 | -44.4 | 6 -2 | 248.2 | 586.5 |
| 6 -4 | 2594 | 977.6 | 6 6 | -582.3 | -2019.5 |
| | | | 6 -6 | -2071 | 692 |

The magnetic dipole hyperfine interaction (8) was written as the scalar product of the nuclear spin moment $\boldsymbol{I}$ and the vector $A_{\mathrm{HF}}\boldsymbol{N}$, which defines the effective magnetic field created by the 4*f* electronic shell at the nucleus [67]. Here, $A_{\mathrm{HF}} = 2\mu_B \gamma_{\mathrm{Ho}} \hbar \langle r^{-3} \rangle_{4f}$ is the hyperfine coupling constant, $\gamma_{\mathrm{Ho}} = 2\pi \cdot 8.98$ MHz/T and $\langle r^{-3} \rangle_{4f} = 9.7$ at. u. being the nuclear gyromagnetic ratio [74] and the inverse third power of the 4*f* electron radius averaged over the electron density distribution [67], respectively. Dimensionless components of the vector $\boldsymbol{N}$ are the following electronic operators:

$$N_x = \sum_k [r_\perp l_{x,k} + (3O_{2,k}^2 - O_{2,k}^0) s_{x,k} / 2 + 3O_{2,k}^{-2} s_{y,k} / 2 + 3O_{2,k}^1 s_{z,k}], \quad (19)$$

$$N_y = \sum_k [r_\perp l_{y,k} + 3O_{2,k}^{-2} s_{x,k} / 2 - (3O_{2,k}^2 + O_{2,k}^0) s_{y,k} / 2 + 3O_{2,k}^{-1} s_{z,k}], \quad (20)$$

$$N_z = \sum_k [r_\| l_{z,k} + O_{2,k}^0 s_{z,k} + 3O_{2,k}^1 s_{x,k} + 3O_{2,k}^{-1} s_{y,k}]. \quad (21)$$

The first term in the operator (9) defines interaction of the quadrupole moment $Q = 2.4 \cdot 10^{-28}$ m$^2$ of a holmium nucleus at the origin of the coordinate frame with the electric field gradient of the ionic crystal lattice ($\boldsymbol{R}_L$ is the radius-vector of a lattice ion with the charge $eq_L$), the second term corresponds to interaction with the $4f^{10}$ electronic shell of the $Ho^{3+}$ ion, $I_0=3I_z^2-I(I+1)$, $I_2=I_x^2-I_y^2$, $I_{-2}=I_xI_y+I_yI_x$, $I_1=I_xI_z+I_zI_x$, $I_{-1}=I_yI_z+I_zI_y$; $R_Q=0.1$ and $\gamma_\infty=-80$ are shielding and antishielding Sternheimer factors [75].

To explore optical transitions within the 4*f* configuration in RE ions, it is necessary to take into account additional interactions of 4*f* electrons with magnetic and electric fields of the electromagnetic radiation and the interaction with the odd crystal field. These interactions are considered as a perturbation. The operator of the effective even electric dipole moment of a RE ion at the site with the $S_4$ symmetry has components

$$\begin{aligned} D_x = \sum_k [ & b_2^1 O_{2,k}^1 + b_2^{-1} O_{2,k}^{-1} + b_4^1 O_{4,k}^1 + b_4^{-1} O_{4,k}^{-1} + b_4^3 O_{4,k}^3 + b_4^{-3} O_{4,k}^{-3} \\ & + b_6^1 O_{6,k}^1 + b_6^{-1} O_{6,k}^{-1} + b_6^3 O_{6,k}^3 + b_6^{-3} O_{6,k}^{-3} + b_6^5 O_{6,k}^5 + b_6^{-5} O_{6,k}^{-5}], \end{aligned} \quad (22)$$

$$\begin{aligned} D_y = \sum_k [ & b_2^{-1} O_{2,k}^1 - b_2^1 O_{2,k}^{-1} + b_4^{-1} O_{4,k}^1 - b_4^1 O_{4,k}^{-1} - b_4^{-3} O_{4,k}^3 + b_4^3 O_{4,k}^{-3} \\ & + b_6^{-1} O_{6,k}^1 - b_6^1 O_{6,k}^{-1} - b_6^{-3} O_{6,k}^3 + b_6^3 O_{6,k}^{-3} + b_6^{-5} O_{6,k}^5 - b_6^5 O_{6,k}^{-5}], \end{aligned} \quad (23)$$

$$\begin{aligned} D_z = \sum_k [ & b_2^2 O_{2,k}^2 + b_2^{-2} O_{2,k}^{-2} + b_4^2 O_{4,k}^2 + b_4^{-2} O_{4,k}^{-2} \\ & + b_6^2 O_{6,k}^2 + b_6^{-2} O_{6,k}^{-2} + b_6^6 O_{6,k}^6 + b_6^{-6} O_{6,k}^{-6}]. \end{aligned} \quad (24)$$

In the present work, in Eqs. (22-24) we used parameters $b_p^q$ calculated for impurity $Er^{3+}$ ions in the isostructural compound $CaMoO_4$ (see Table 4) [76].

Table 4. Parameters of the effective electric dipole moment ($10^{-4}$ $e \cdot$ nm).

| $p$ $q$ | $b_p^q$ | $p$ $q$ | $b_p^q$ |
|---|---|---|---|
| 2 1 | 6.2 | 2 2 | 1.2 |
| 2 –1 | –7.7 | 2 -2 | -0.12 |
| 4 1 | -8.5 | 4 2 | 3.12 |
| 4 -1 | 18.1 | 4 -2 | 0.51 |
| 6 1 | 2.7 | 6 2 | -2.0 |
| 6 –1 | –6.2 | 6 -2 | 13.1 |
| 4 3 | -20.8 | 6 6 | -1.35 |
| 4 -3 | -25.6 | 6 -6 | -0.08 |
| 6 3 | –6.5 | | |
| 6 –3 | 18.9 | | |
| 6 5 | –28.8 | | |
| 6 –5 | 4.6 | | |

The matrices of all electronic operators introduced above were calculated in the total basis of $C_{14}^{10}$=1001 Slater determinants of the $4f^{10}$ configuration. CF energy levels and corresponding electronic wave functions, as well as HFS of CF singlets and doublets were obtained using the two-step numerical diagonalizations. At the first step, the operator $H_0$+$H_{CF}$ was diagonalized, and the total Hamiltonian (4) was projected on the obtained space of wave functions. At the second step, the electron-nuclear Hamiltonian (4) was diagonalized in the truncated Hilbert space of 608 states corresponding to lower electronic multiplets $^5I_J$ ($J$=4 - 8) and $^5F_5$.

Modeling of the registered luminescence spectra involved variations of CF parameters, the orbital reduction factor $r_{||}$, and the width $\xi$ of the distribution function of random deformations. The values of CF parameters obtained from the fitting of CF energies are presented in Table 2.

The calculated final energies of CF sublevels of six lower multiplets and total widths of HFS of non-Kramers doublets are compared with the results of measurements in Table 1 and Table S3 in the Supplemental Material [58]. Differences between the CF parameters used in the present work and parameters presented earlier in the literature [56,72] are relatively small, however, the agreement between the simulated CF structures of the considered multiplets and the experimental data is markedly improved. Some contradictions between locations of the quasi-degenerate sublevels of different symmetry relative to each other are caused by neglecting the hyperfine interaction that shifts centers of gravity of HFS of singlets and doublets by about one $cm^{-1}$. The orbital reduction factor $r_{||}$ =0.981 was determined from a comparison of the calculated and measured $g$-factors $g_{zz} = g_{||}$ of the ground doublet $\Gamma_{34}{}^1({}^5I_8)$.

Envelopes of spectral lines in unpolarized luminescence spectra corresponding to transitions $\Gamma i \rightarrow \Gamma' f$ between CF sublevels of different multiplets at temperature $T$ were simulated using the following intensity distribution for the fixed external magnetic field $\boldsymbol{B}$ and deformation tensor components $e(\Gamma\lambda)$:

$$I_{\Gamma i,\Gamma' f}(\boldsymbol{B},\boldsymbol{e}\,|\,\omega) \propto \sum_{j\in\Gamma i}\sum_{j'\in\Gamma' f} S_{jj'}(\boldsymbol{B},\boldsymbol{e}) \exp[-(E_j - E_{j0})/k_B T]\Delta_{\Gamma i\Gamma' f}[(E_j - E_{j'} - \hbar\omega)^2 + \Delta^2_{\Gamma i\Gamma' f}]^{-1}. \quad (25)$$

In Eq. (25), summations are taken over hyperfine sublevels $j$ and $j'$ of the initial $\Gamma i$ and final $\Gamma' f$ CF energy levels of a considered transition, $E_{j0}$ is the energy of the lowest hyperfine sublevel of the initial CF state, $S_{jj'} = n^3 \sum_\alpha |< j'|M_\alpha|j>|^2 + \chi \sum_\alpha |< j'|D_\alpha|j>|^2$ is the sum of the magnetic dipole and electric dipole line strengths. The factor $\chi=[(n^2+2)/3n]^2$, where $n$ is the refractive index, is introduced to account for the local electric field correction. For $CaWO_4$, $n$=1.94 and, thus, $\chi\approx 1$. The Boltzmann distribution of populations of hyperfine sublevels of the initial CF state is assumed, and the Lorentz lineshape with the halfwidth $\Delta_{\Gamma i\Gamma' f}$ of individual transitions between hyperfine sublevels is used.

The final lineshapes were obtained by averaging envelopes (25) with the distribution function (13) of random deformations, where the width $\xi=3\cdot 10^{-6}$ was determined using the measured deformation splitting $\Delta$=0.025 $cm^{-1}$ of the ground doublet (see Fig. 6b). The calculated profiles of HFS are compared with the measured spectra in Figs. 2 and 6.

## V. CONCLUSIONS

We have measured the photoluminescence spectra of $CaWO_4:Ho^{3+}$(0.01 at. %) single crystal. The PL spectra were recorded in the spectral range 5000 – 13,000 $cm^{-1}$ with high spectral resolution (0.02 $cm^{-1}$), under excitation by the light of a diode laser ($\lambda_{ex}$ = 641.6 nm). All observed lines could be identified as associated with $Ho^{3+}$ ions replacing $Ca^{2+}$ ions at the $S_4$ tetragonal symmetry sites with non-local charge compensation. Well-resolved hyperfine structure was observed for the first time in the luminescence spectra of $CaWO_4:Ho^{3+}$. The lines of transitions involving the CF levels of the holmium ${}^5I_8$ ground multiplet have the same structure in the luminescence and absorption spectra and show no traces of deformation splitting. For the lines with HFS, the deformation splitting caused by the interaction of degenerate electronic levels with random crystal-lattice strains manifests itself in an increase in the central interval of the HFS. In the *absorption* spectra of $CaWO_4:Ho^{3+}$, the deformation splitting was absent for holmium concentrations less than 0.5 at. % [56]. In the *luminescence* spectra, transitions between excited multiplets are observed, which cannot be observed in the absorption spectra. In the luminescence spectrum of $CaWO_4$:$Ho^{3+}$(0.01 at. %), we have found a pronounced increase in the central interval of the HFS for the line at 6740.3 $cm^{-1}$ belonging to the $^5F_5 \rightarrow {}^5I_6$ spectral manifold. This PL line, which falls within the S-band transparency window of optical fibers, can serve as a sensitive indicator of internal lattice strains for crystal quality control.

The PL spectra of $CaWO_4$:$Ho^{3+}$ provide the possibility of implementing a thermometer for cryogenic temperatures. Here, a Boltzmann ratiometric luminescence thermometer based on the lines 8640.3 $cm^{-1}$ and 8582.7 $cm^{-1}$ in the PL spectra of $CaWO_4$:$Ho^{3+}$(0.5 at. %) is realized. It has maximum absolute thermal sensitivity at the temperature $T_m$ = 21.5 K, the relative sensitivity at this temperature is 10 % $K^{-1}$.

Magnetic field dependent (0 – 270 mT) spectra were registered at 15 K for magnetic field directed along the *c* axis of this tetragonal Scheelite-type crystal and the *g* factors were determined. We show that using the PL line at 5160.5 $cm^{-1}$ with $g$ = 25 corresponding to the $^5I_7\,\Gamma_{34}$ (5160.5) $\rightarrow$ $^5I_8\,\Gamma_{34}$ (0) transition it is possible to detect variations of a magnetic field as small as several tenths of mT.

Using the data on energies of crystal-field levels specified on the basis of analysing high-resolution PL spectra and the values of determined *g* factors, we have obtained a refined set of the crystal-field parameters for the $Ho^{3+}$ ion in $CaWO_4$:$Ho^{3+}$ crystal. Though the differences between the CF parameters used in the present work and parameters presented earlier in the literature was relatively small, the agreement between the calculated CF structures of holmium multiplets and the experimental data was significantly improved. Modeling of the observed PL spectra was performed taking into account hyperfine and deformation interactions and using the distribution function of random strains induced by point lattice defects in an elastically anisotropic $CaWO_4$ crystal. The simulation has confirmed an outstanding sensitivity of the PL line 6740.3 $cm^{-1}$ [$^5F_5\,\Gamma_2$ (154 10.1$cm^{-1}$) $\rightarrow$ $^5I_6\,\Gamma_{34}$ (8669.8 $cm^{-1}$) transition] to lattice deformations.

**Acknowledgments**

The authors thank K. N. Boldyrev, N. Yu. Boldyrev, and E. S. Sektarov for their help in some measurements. Financial support of the Russian Science Foundation under Grant No 23-12-00047 is acknowledged.

## Supplemental Material for

# High-resolution luminescence spectroscopy of $CaWO_4:Ho^{3+}$: Sensitive detection of internal strains, temperature, and magnetic field

M. N. Popova[1,*], M. Diab[1,2], N.N. Kuzmin[1], K.A. Subbotin[3], A.I.Titov[3], B. Z. Malkin[4]

[1]*Institute of Spectroscopy, Russian Academy of Sciences, Troitsk, Moscow, 108840, Russia*
[2]*Moscow Institute of Physics and Technology (National Research University), Dolgoprudny, Moscow region, 141700, Russia*
[3]*Prokhorov General Physics Institute of the Russian Academy of Sciences, Moscow, 119991, Russia*
[4]*Kazan Federal University, Kazan, 420008, Russia*

**1.** Selection rules, number of crystal-field levels in *J* multiplets (**Tables S1, S2**)

**2**. Energy levels, hyperfine structure, and *g* factors in $CaWO_4:Ho^{3+}$. (**Table S3**)

**3.** Photoluminescence spectra of $CaWO_4:Ho^{3+}$ in zero magnetic field and spectral line identification (**Figs. S1 – S3; Tables S4 – S6**)

**4**. Photoluminescence spectra of $CaWO_4:Ho^{3+}$ in a magnetic field B||*c*. Determination of *g*- factors (**Figs. S4 – S8**).

**5.** Free-ion parameters for $Ho^{3+}$ in $CaWO_4:Ho^{3+}$

## 1. Selection rules, number of crystal-field levels in *J* multiplets

**Table S1.** Selection rules. $d_k$ ($m_k$), k = x, y, z, denote the allowed components of the ED (MD) transitions. For convenience, also polarizations are indicated, e.g. $\sigma_e \pi_m$ means that the transition is ED allowed in the σ polarization (**k**⊥*c*, **E**⊥*c*) and MD allowed in the π polarization (**k**⊥*c*, **E**||*c*).

| $S_4$ | $\Gamma_1$ | $\Gamma_2$ | $\Gamma_3$ | $\Gamma_4$ |
|---|---|---|---|---|
| $\Gamma_1$ | $m_z$ | $d_z$ | $d_{x-iy}$, $m_{x+i,y}$ | $d_{x+iy}$, $m_{x-i,y}$ |
| | $\sigma_m$ | $\pi_e$ | $\alpha_e$ $\sigma_e$ $\alpha_m$ $\pi_m$ | |
| $\Gamma_2$ | | $m_z$ | $d_{x+iy}$, $m_{x-i,y}$ | $d_{x-iy}$, $m_{x+i,y}$ |
| | | $\sigma_m$ | $\alpha_e$ $\sigma_e$ $\alpha_m$ $\pi_m$ | |
| $\Gamma_3$ | | | $m_z$ | $d_z$ |
| | | | $\sigma_m$ | $\pi_e$ |
| $\Gamma_4$ | | | $d_z$ | $m_z$ |
| | | | $\pi_e$ | $\sigma_m$ |

**Table S2.** Number of CF levels originated from a level with a given value of the total moment *J* of a free ion with even number of electrons, placed into the $S_4$-symmetry position in a crystal, and corresponding irreducible representations of CF levels.

| ***J*** | **Number of CF levels (*S*4)** | **Irreducible representations** |
|---|---|---|
| 8 | 13 | $5\Gamma_1+4\Gamma_2+4\Gamma_{34}$ |
| 7 | 11 | $3\Gamma_1+4\Gamma_2+4\Gamma_{34}$ |
| 6 | 10 | $3\Gamma_1+4\Gamma_2+3\Gamma_{34}$ |
| 5 | 8 | $3\Gamma_1+2\Gamma_2+3\Gamma_{34}$ |

## 2. Energy levels of $Ho^{3+}$ in $CaWO_4:Ho^{3+}$

**Table S3**. Irreducible representations IR, crystal-field energies *E* ($cm^{-1}$), and total hyperfine splittings $\Delta_{HF}$ ($cm^{-1}$) and *g* factors of $\Gamma_{34}$ doublets of $Ho^{3+}$ in tetragonal sites of $CaWO_4:Ho^{3+}$ [S1] and present work. Asterisks mark the data from Ref. [S2].

| Level | | IR | $E_{exp}$ | $E_{theor}$ | $\Delta_{HF}$ exp | $\Delta_{HF}$ theor | $g_{\|}$ exp | $g_{\|}$ theor |
|---|---|---|---|---|---|---|---|---|
| $^5F_5$ | 8 | $\Gamma_{34}^3$ | 15585 | 15590 | - | 0.67 | - | 11.29 |
| | 7 | $\Gamma_1^3$ | 15568 | 15566 | | | | |
| | 6 | $\Gamma_2^2$ | 15551.8 | 15558 | | | | |
| | 5 | $\Gamma_{34}^2$ | 15544 | 15550 | - | 0.07 | - | 1.27 |
| | 4 | $\Gamma_1^2$ | 15478.4 | 15482 | | | | |
| | 3 | $\Gamma_1^1$ | 15437.7 | 15443 | | | | |
| | 2 | $\Gamma_{34}^1$ | 15427.5 | 15425 | - | 0.26 | - | 4.56 |
| | 1 | $\Gamma_2^1$ | 15410.1 | 15421 | | | | |
| | | | | | | | | |
| $^5I_4$ | 7 | $\Gamma_1^3$ | 13499* | 13502 | | | | |
| | 6 | $\Gamma_{34}^2$ | 13388* | 13390 | - | 0.1851 | - | 0.872 |
| | 5 | $\Gamma_2^2$ | - | 13319 | | | | |
| | 4 | $\Gamma_1^2$ | 13308* | 13313 | | | | |
| | 3 | $\Gamma_2^1$ | - | 13311 | | | | |
| | 2 | $\Gamma_{34}^1$ | 13253* | 13248 | - | 0.8268 | - | 3.202 |
| | 1 | $\Gamma_1^1$ | 13180* | 13176 | | | | |
| | | | | | | | | |
| $^5I_5$ | 8 | $\Gamma_2^2$ | 11313* | 11306 | | | | |
| | 7 | $\Gamma_{34}^3$ | 11310* | 11302 | - | 0.06 | - | 0.23 |
| | 6 | $\Gamma_1^3$ | 11288* | 11280 | | | | |
| | 5 | $\Gamma_{34}^2$ | 11230.4 | 11218 | - | 0.54 | - | 3.67 |
| | 4 | $\Gamma_1^2$ | 11227.3 | 11224 | | | | |
| | 3 | $\Gamma_2^1$ | 11221.2 | 11219 | | | | |
| | 2 | $\Gamma_1^1$ | 11215* | 11212 | | | | |
| | 1 | $\Gamma_{34}^1$ | 11209.8 | 11210 | 1.33 | 1.28 | 9.9 | 8.69 |
| | | | | | | | | |
| $^5I_6$ | 10 | $\Gamma_2^4$ | 8773.9 | 8772.0 | | | | |
| | 9 | $\Gamma_{34}^3$ | 8763.2 | 8761.5 | - | 0.08 | - | 0.53 |
| | 8 | $\Gamma_1^3$ | 8751.1 | 8750.0 | | | | |
| | 7 | $\Gamma_2^3$ | 8680.5 | 8676.2 | | | | |
| | 6 | $\Gamma_1^2$ | 8675.1 | 8670.4 | | | | |
| | 5 | $\Gamma_{34}^2$ | 8669.8 | 8662.9 | 0.37 | 0.32 | 3.4 | 3.03 |
| | 4 | $\Gamma_{34}^1$ | 8658.2 | 8655.1 | 1.07 | 1.03 | 9.9 | 9.93 |
| | 3 | $\Gamma_2^2$ | 8657.4 | 8660.5 | | | | |
| | 2 | $\Gamma_1^1$ | 8650.6 | 8650.7 | | | | |
| | 1 | $\Gamma_2^1$ | 8650.5 | 8650.5 | | | | |
| | | | | | | | | |
| $^5I_7$ | 11 | $\Gamma_1^3$ | 5275* | 5275.7 | | | | |
| | 10 | $\Gamma_{34}^2$ | 5274 | 5274.1 | - | 0.23 | - | 0.69 |
| | 9 | $\Gamma_2^4$ | 5269 | 5271.3 | | | | |
| | 8 | $\Gamma_2^3$ | 5218.4 | 5217.9 | | | | |
| | 7 | $\Gamma_{34}^3$ | 5207.5 | 5210.6 | - | 0.54 | - | 5.82 |
| | 6 | $\Gamma_1^2$ | 5189.0 | 5188.7 | | | | |
| | 5 | $\Gamma_{34}^2$ | 5160.5 | 5166.5 | 1.05 | 1.04 | 11.1 | 11.86 |
| | 4 | $\Gamma_1^1$ | 5154.4 | 5153.5 | | | | |
| | 3 | $\Gamma_2^2$ | 5144 | 5145.5 | | | | |
| | 2 | $\Gamma_{34}^1$ | 5143.9 | 5143.7 | 0.69 | 0.68 | 6.4 | 7.76 |
| | 1 | $\Gamma_2^1$ | 5137.2 | 5138.1 | | | | |
| | | | | | | | | |
| | 13 | $\Gamma_2^4$ | 325* | 294.8 | | | | |
| $^5I_8$ | 12 | $\Gamma_1^5$ | 297 | 277.1 | | | | |
| | 11 | $\Gamma_{34}^4$ | 282 | 274.9 | - | 0.78 | - | 9.94 |
| | 10 | $\Gamma_{34}^3$ | 262* | 265.6 | - | 0.30 | - | 4.08 |
| | 9 | $\Gamma_1^4$ | 254* | 259.3 | | | | |
| | 8 | $\Gamma_2^3$ | 248* | 253.3 | | | | |
| | 7 | $\Gamma_1^3$ | 197.5 | 201.9 | | | | |
| | 6 | $\Gamma_{34}^2$ | 67.8 | 71.0 | 0.70 | 0.75 | - | 9.73 |
| | 5 | $\Gamma_1^2$ | 46.2 | 53.1 | | | | |
| | 4 | $\Gamma_1^1$ | 40.5 | 46.7 | | | | |
| | 3 | $\Gamma_2^2$ | 22.2 | 20.8 | | | | |
| | 2 | $\Gamma_2^1$ | 9.2 | 9.2 | | | | |
| | 1 | $\Gamma_{34}^1$ | 0 | 0 | 1.05 | 1.05 | 13.9 | 13.69 |

## 3. Photoluminescence spectra of $CaWO_4:Ho^{3+}$ in zero magnetic field

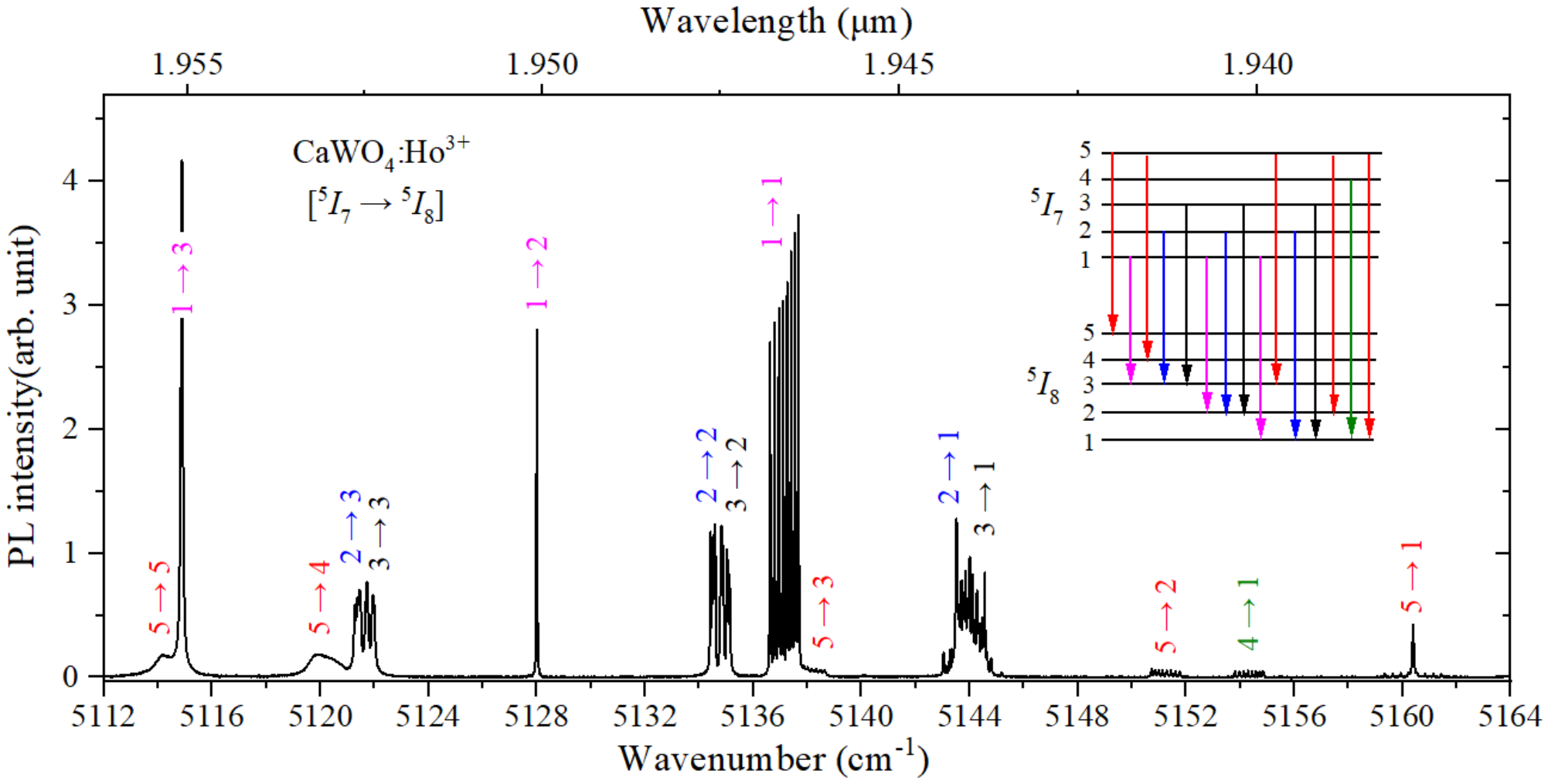

**Fig. S1.** PL spectrum of $CaWO_4:Ho^{3+}$ (0.01 at. %) in the region of the transition $^5I_7 \rightarrow {}^5I_8$. $T = 15$ K; $\lambda_{ex} = 641.6$ nm.

**Table S4**. Frequencies (cm$^{-1}$) of transitions in the luminescence channel $^5I_7 \rightarrow {}^5I_8$ averaged over HFS components of the corresponding spectral line.

| $^5I_8$ \ $^5I_7$ | | 1 | 2 | 3 | 4 | 5 |
|---|---|---|---|---|---|---|
| | | 5137.2 ($\Gamma_2^1$) | 5143.9 ($\Gamma_{34}^1$) | 5144 ($\Gamma_2^2$) | 5154.4 ($\Gamma_1^1$) | 5160.5 ($\Gamma_{34}^2$) |
| 5 | 46.2 ($\Gamma_1^2$) | 5091 | 5097.7 | 5097.8 | 5108.2 | 5114.3 |
| 4 | 40.5 ( $\Gamma_1^1$) | 5096.7 | 5103.4 | 5103.5 | 5113.9 | 5120 |
| 3 | 22.2 ($\Gamma_2^2$) | 5115 | 5121.7 | 5121.8 | 5132.2 | 5138.3 |
| 2 | 9.2 ($\Gamma_2^1$) | 5128 | 5134.7 | 5134.8 | 5145.2 | 5151.3 |
| 1 | 0 ($\Gamma_{34}^1$) | 5137.2 | 5143.5 | 5144 | 5154.4 | 5160.5 |

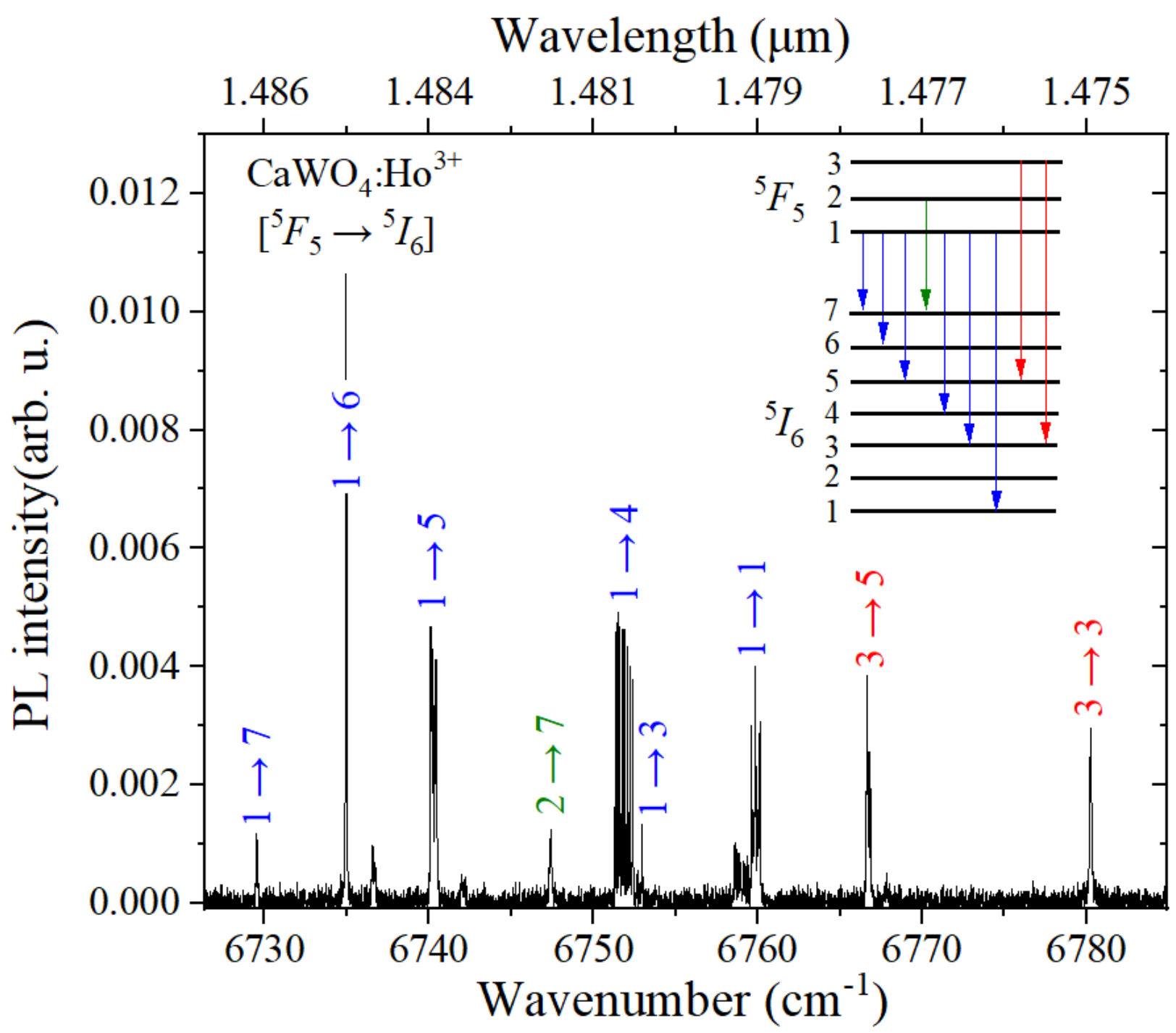

**Fig. S2.** PL spectrum of $CaWO_4:Ho^{3+}$ (0.01 at. %) in the region of the transition $^5F_5 \rightarrow {}^5I_6$. $T$ = 15 K; $\lambda_{ex}$ = 641.6 nm.

**Table S5**. . Frequencies ($cm^{-1}$) of transitions in the luminescence channel $^5F_5 \rightarrow {}^5I_6$ averaged over HFS components of the corresponding spectral line.

| $^5I_6$ \ $^5F_5$ | | 1 | 2 | 3 | 4 |
|---|---|---|---|---|---|
| | | 15410.1 ($\Gamma_2^1$) | 15427.5 ($\Gamma_{34}^1$) | 15437.7 ($\Gamma_1^1$) | 15478.4 ($\Gamma_1^2$) |
| 6 | 8675.1 ($\Gamma_1^2$) | 6735 | 6752.4 | 6762.6 | 6803.3 |
| 5 | 8669.8 ($\Gamma_{34}^2$) | 6740.3 | 6757.7 | 6767.9 | 6808.6 |
| 4 | 8658.2 ($\Gamma_{34}^1$) | 6751.9 | 6769.3 | 6779.5 | 6820.2 |
| 3 | 8657.4 ($\Gamma_2^2$) | 6752.7 | 6770.1 | 6780.3 | 6821 |
| 2 | 8650.6 ($\Gamma_1^1$) | 6759.5 | 6776.9 | 6787.1 | 6827.8 |
| 1 | 8650.5 ($\Gamma_2^1$) | 6759.6 | 6777 | 6787.2 | 6827.9 |

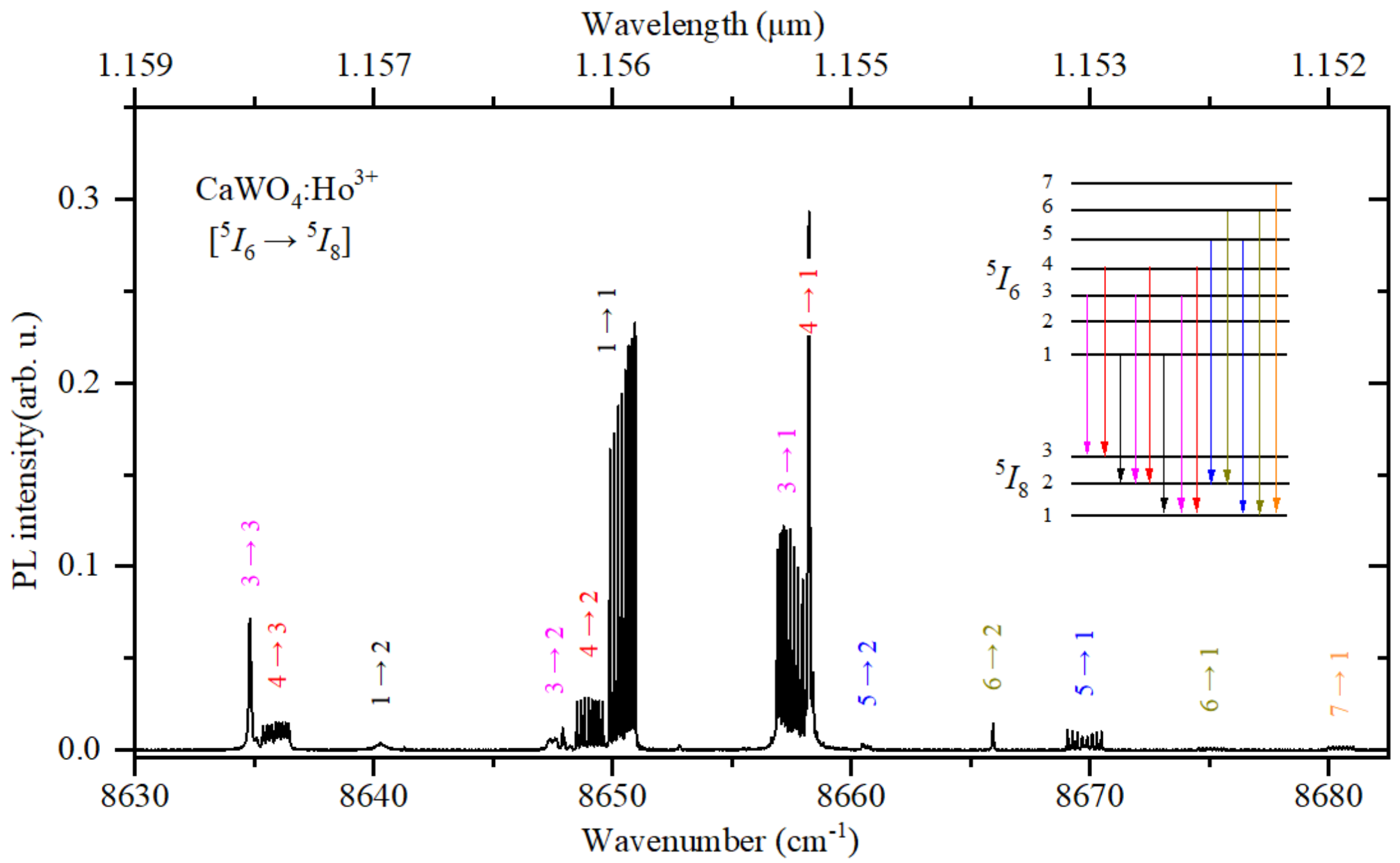

**Fig. S3.** PL spectrum of CaWO4:Ho$^{3+}$ (0.01 at. %) in the region of the transition $^5I_6 \rightarrow {}^5I_8$. $T = 15$ K; $\lambda_{ex} = 641.6$ nm.

**Table S6**. Frequencies (cm$^{-1}$) of transitions in the luminescence channel $^5I_6 \rightarrow {}^5I_8$ averaged over HFS components of the corresponding spectral line.

| $^5I_8$ \ $^5I_6$ | | 1 | 2 | 3 | 4 | 5 | 6 | 7 |
|---|---|---|---|---|---|---|---|---|
| | | 8650.5 ($\Gamma_2^1$) | 8650.6 ($\Gamma_1^1$) | 8657.4 ($\Gamma_2^2$) | 8658.2 ($\Gamma_{34}^1$) | 8669.8 ($\Gamma_{34}^2$) | 8675.1 ($\Gamma_1^2$) | 8680.5 ($\Gamma_2^3$) |
| 6 | 67.8 ($\Gamma_{34}^2$) | 8682.7 | 8682.8 | 8589.6 | 8590.4 | 8602 | 8607.3 | 8612.7 |
| 5 | 46.2 ($\Gamma_1^2$) | 8604.3 | 8604.4 | 8611.2 | 8612 | 8623.6 | 8628.9 | 8634.3 |
| 4 | 40.5( $\Gamma_1^1$) | 8610 | 8610.1 | 8616.9 | 8617.7 | 8629.3 | 8634.6 | 8640 |
| 3 | 22.2 ($\Gamma_2^2$) | 8628.3 | 8628.4 | 8635.23 | 8636.03 | 8647.63 | 8652.93 | 8658.33 |
| 2 | 9.2 ($\Gamma_2^1$) | 8641.3 | 8641.4 | 8648.24 | 8649.04 | 8660.64 | 8665.94 | 8671.34 |
| 1 | 0 ($\Gamma_{34}^1$) | 8650.5 | 8650.6 | 8657.4 | 8658.2 | 8669.8 | 8675.1 | 8680.5 |

## 4. Photoluminescence spectra of CaWO4:Ho$^{3+}$ in a magnetic field $B\|c$. Determination of $g$ factors

## $^5I_7 \rightarrow {}^5I_8$

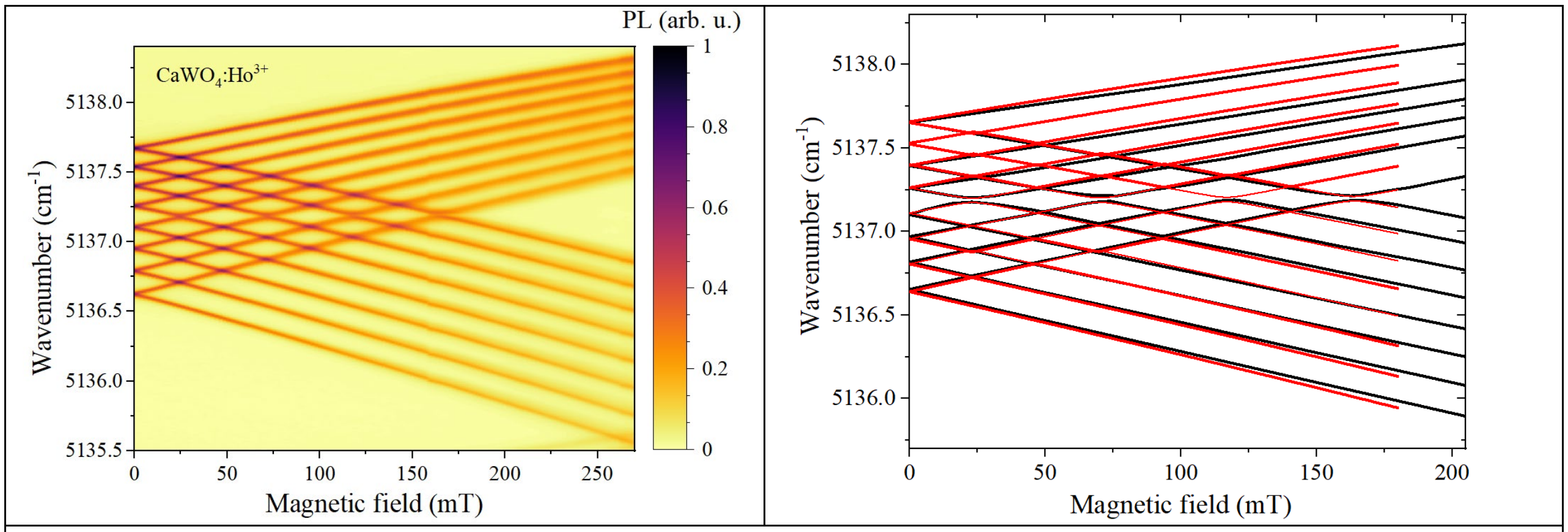

(*a*) $^5I_7\ \Gamma_2$ (5137.2) ⇒ $^5I_8\ \Gamma_{34}$ (0) (*b*)

**Fig. S4.** Luminescence intensity map in the magnetic-field – wavenumber scale for the line 5137.2 cm$^{-1}$ in the spectrum of $CaWO_4:Ho^{3+}$ (0.01 at. %) in a magnetic field **B**||***c***. $T = 15$ K; $\lambda_{ex} = 641.6$ nm. (a) Experiment; (b) Experimental (black) and calculated (red) frequencies in the magnetic-field – wavenumber scale for the same line.

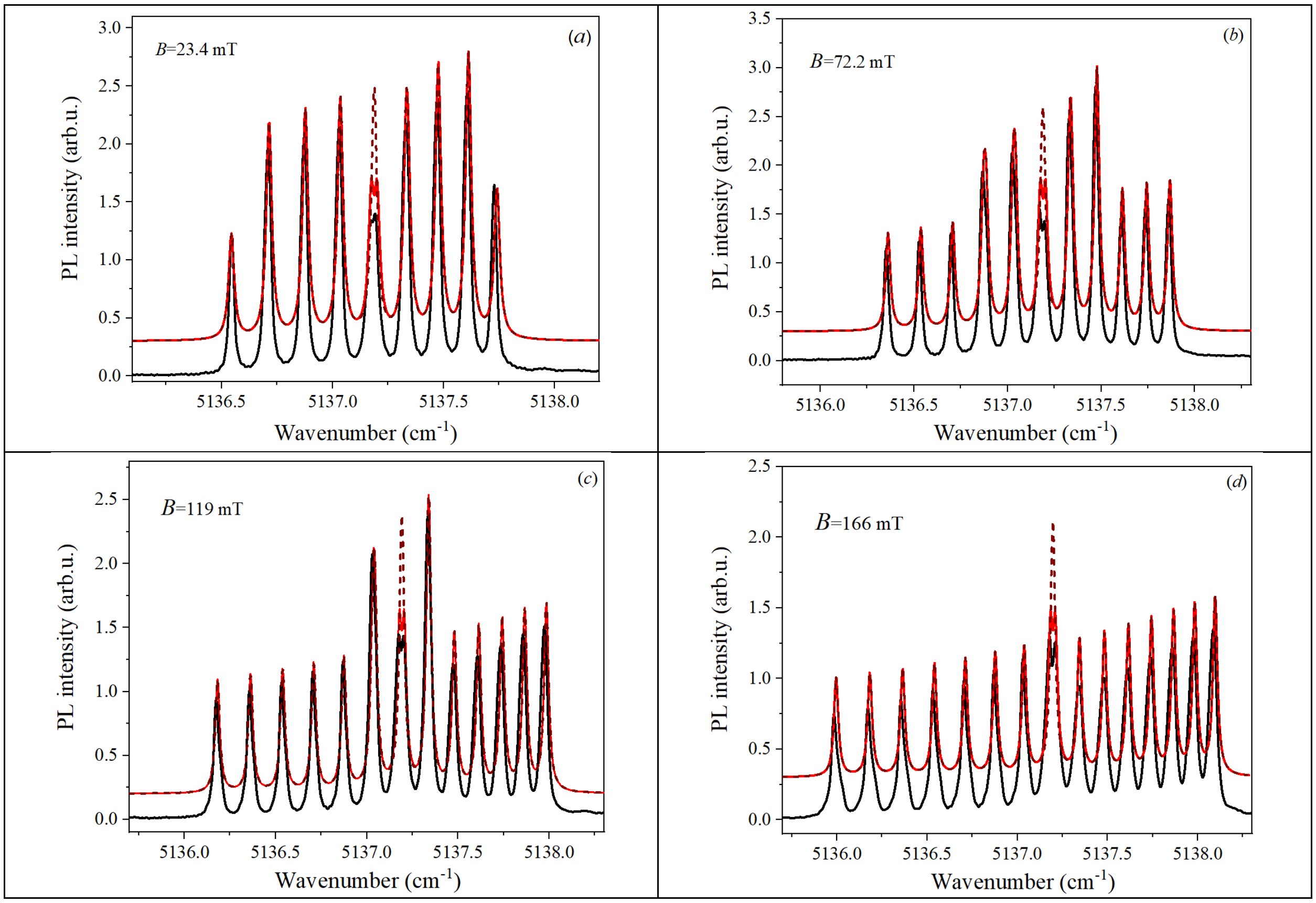

**Fig. S5.** Anticrossings $\Delta m$=0 in the line 5137.2 cm$^{-1}$ corresponding to the singlet-doublet transition [$^5I_7\,\Gamma_2{}^1$ (5137.2) → $^5I_8\,\Gamma_{34}{}^1$ (0)] in magnetic fields $B$=23.4 (*a*), 72 (b), 118 (c) and 166 (d) mT. The registered HFS envelops are shown by black lines, the results of modeling with and without taking into account random strains are shown by the solid red and dash purple lines, respectively. The values of magnetic fields *B* at the anticrossings presented in the figures (*a*-*d*) agree satisfactorily with the corresponding fields of 23.5, 70.9, 117.0, and 165 mT which were determined from EPR spectra in [S1].

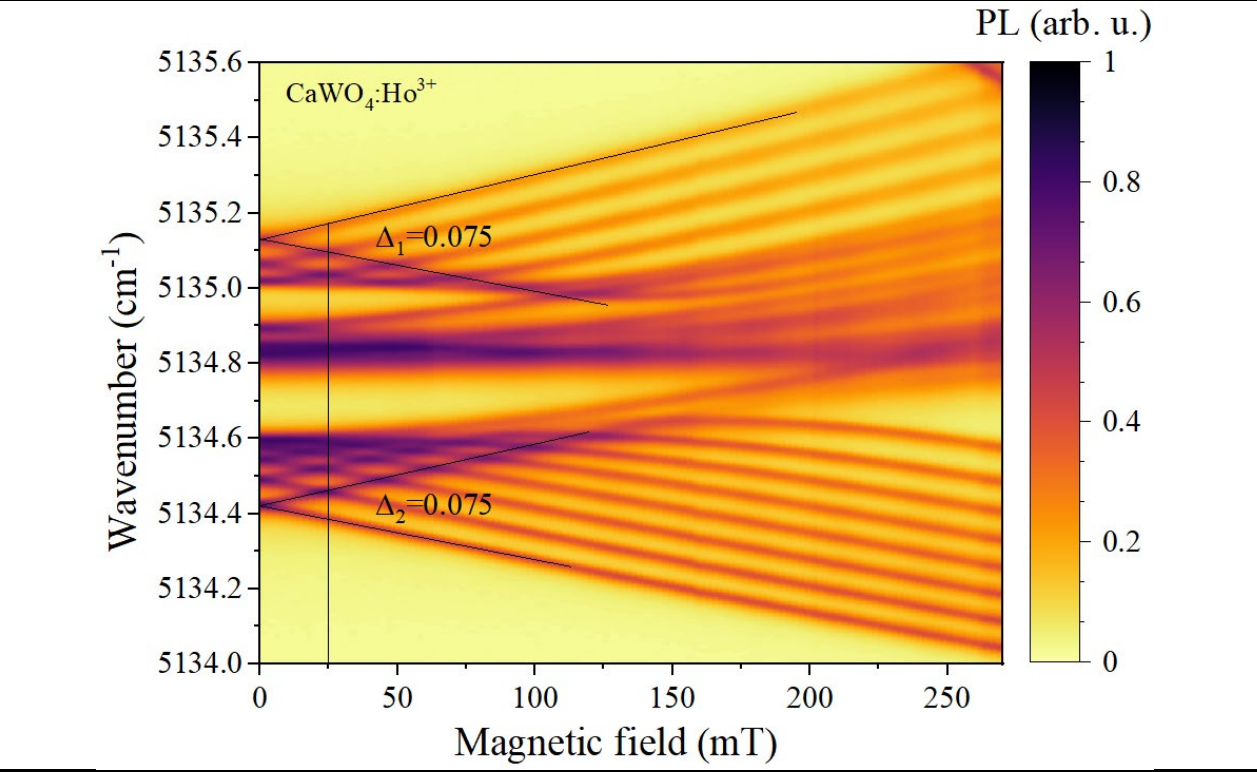

[$^5I_7\,\Gamma_{34}{}^1$ (5143.5) → $^5I_8\,\Gamma_2{}^1$ (9.16)]+
[$^5I_7\,\Gamma_2{}^2$ (5144) → $^5I_8\,\Gamma_2{}^1$ (9.16)].
$\Delta_{HF}$=0.709 cm$^{-1}$  $g_1 = g_2 = \langle g \rangle = 6.4$

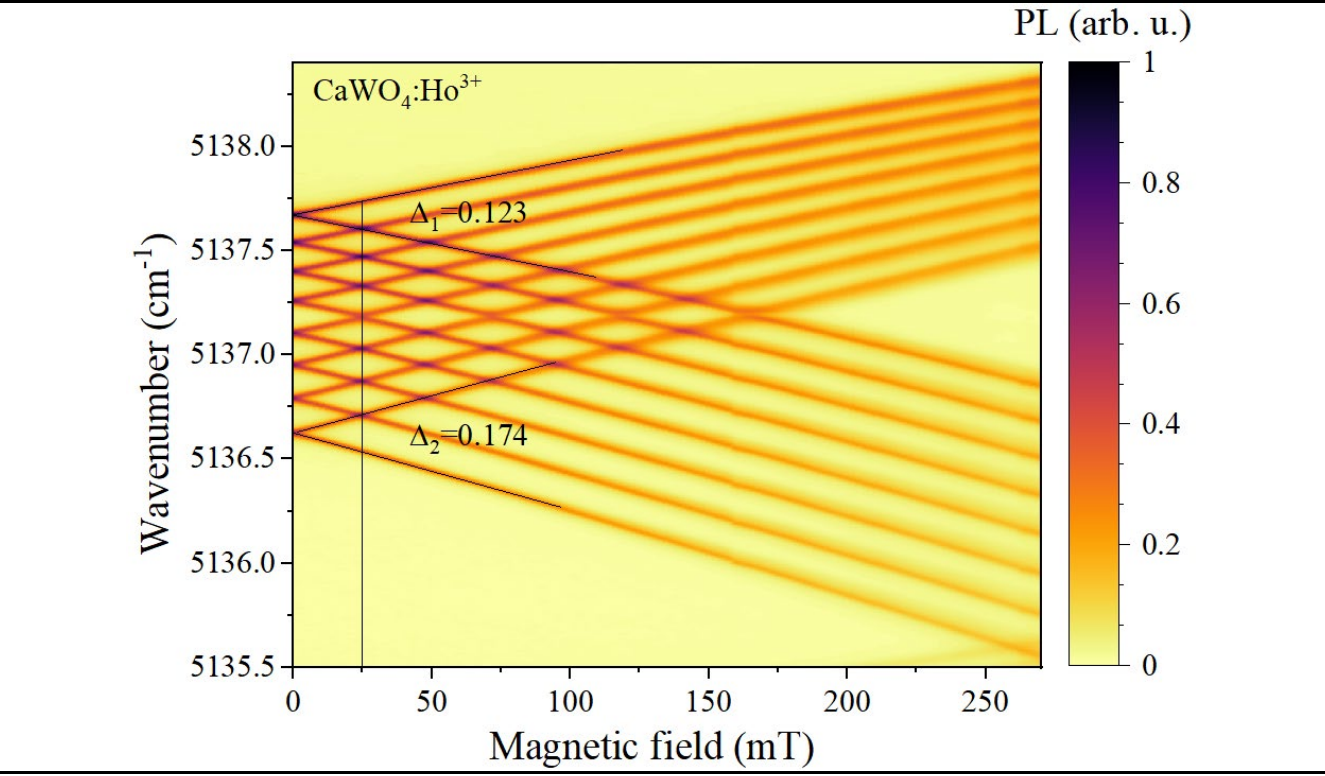

[$^5I_7\,\Gamma_2$ (5137.2) → $^5I_8\,\Gamma_{34}$ (0)]. $\Delta_{HF}$=1.047 cm$^{-1}$
$g_1 = 10.5$, $g_2 = 14.9$; $\langle g \rangle = 12.72$

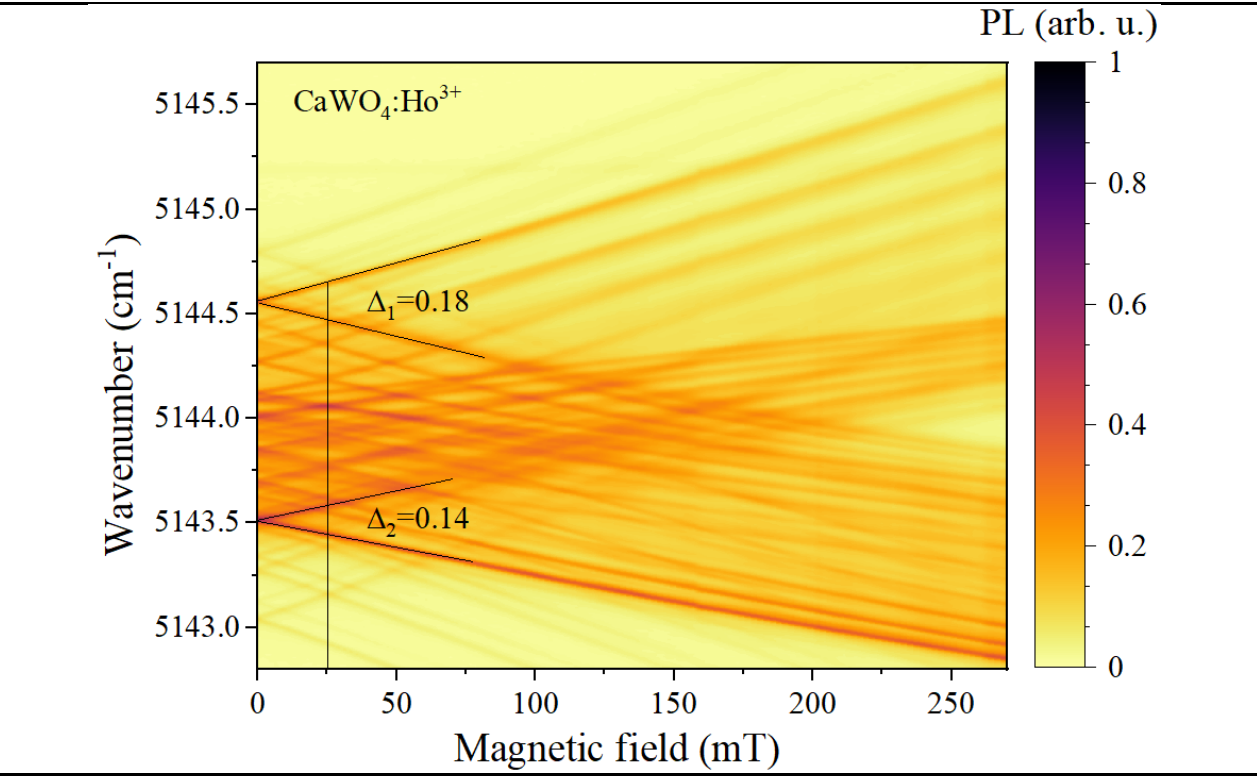

[$^5I_7\,\Gamma_{34}{}^1$ (5143.3) → $^5I_8\,\Gamma_{34}{}^1$ (0)].
$\Delta_{HF}$=1.041 cm$^{-1}$
$g_1 = 15.4$, $g_2 = 12.0$; $\langle g \rangle = 13.7$

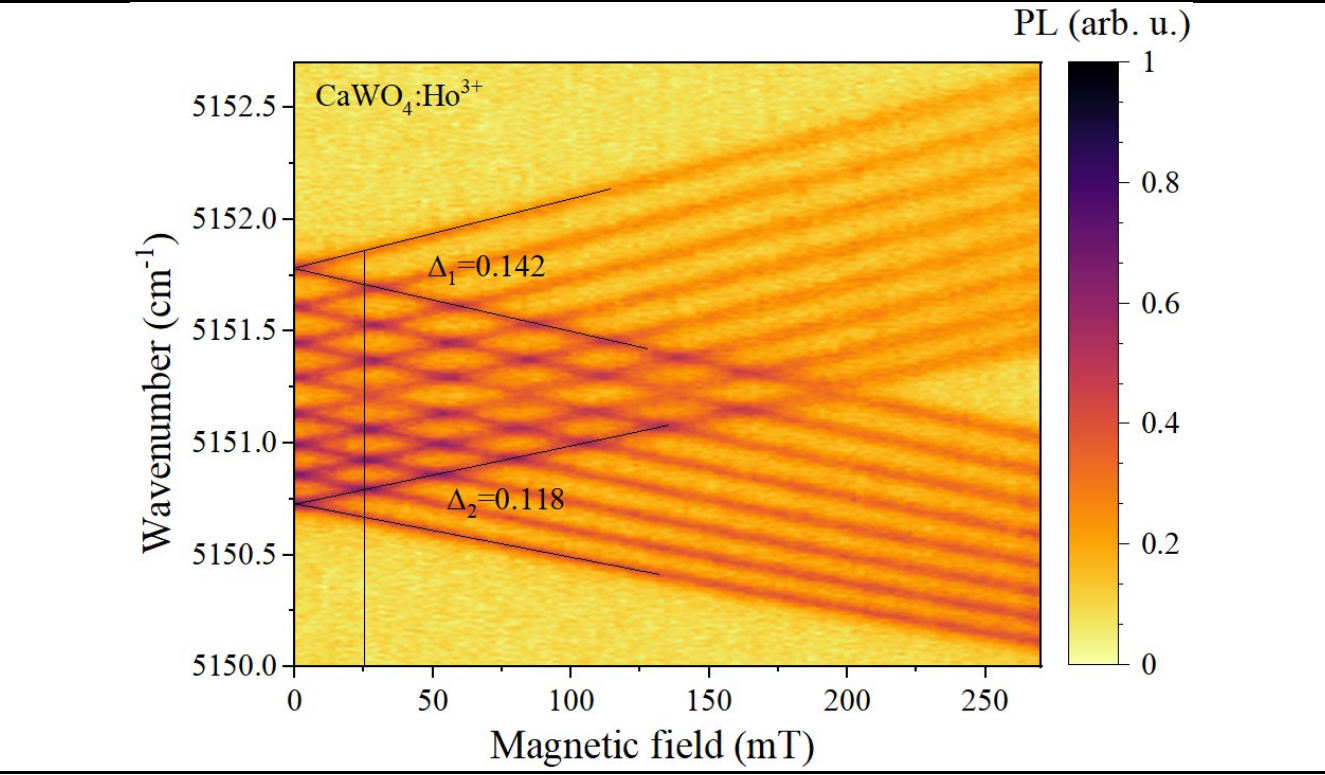

[$^5I_7\,\Gamma_{34}{}^2$ (5160.5) → $^5I_8\,\Gamma_2{}^1$ (9.16)].
$\Delta_{HF}$=1.053 cm$^{-1}$
$g_1 = 12.2$, $g_2 = 10.1$; $\langle g \rangle = 11.14$

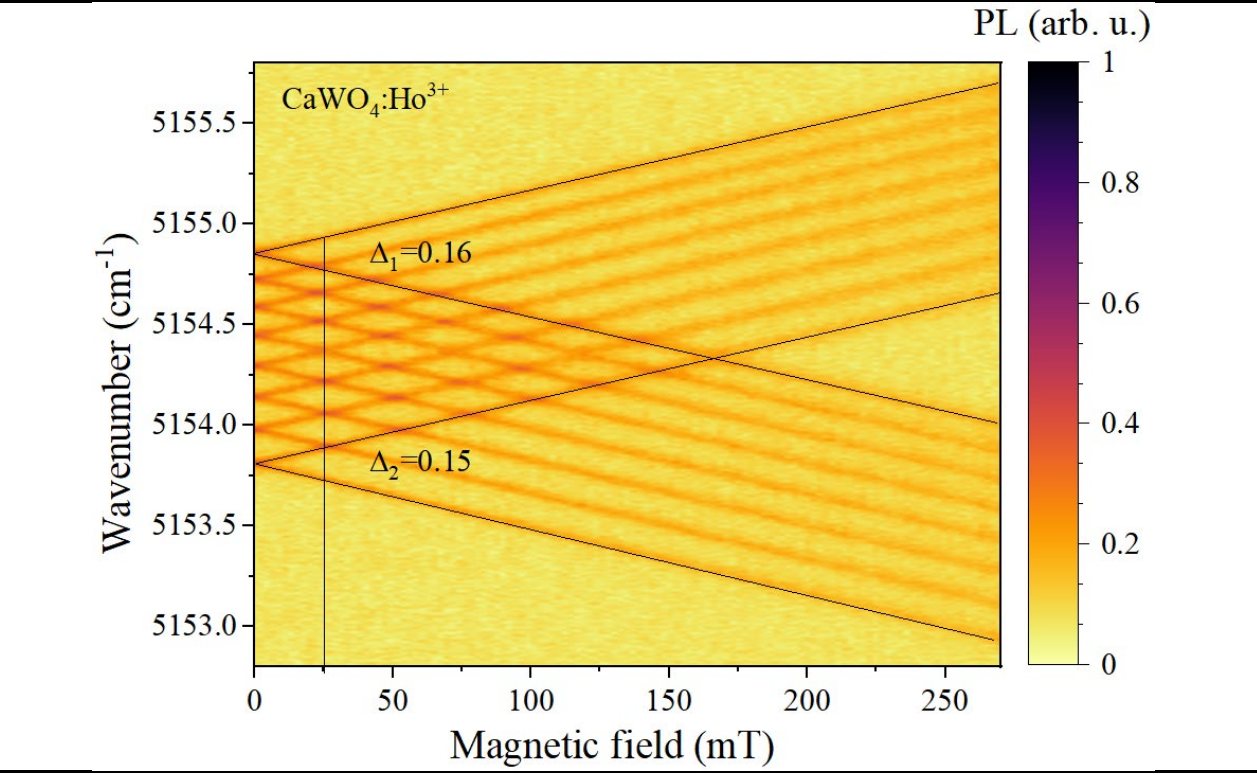

[$^5I_7\,\Gamma_1{}^1$ (5154.4) → $^5I_8\,\Gamma_{34}{}^1$ (0)]. $\Delta_{HF}$=1.05 cm$^{-1}$
$g_1 = 13.71$, $g_2 = 12.85$; $\langle g \rangle = 13.28$

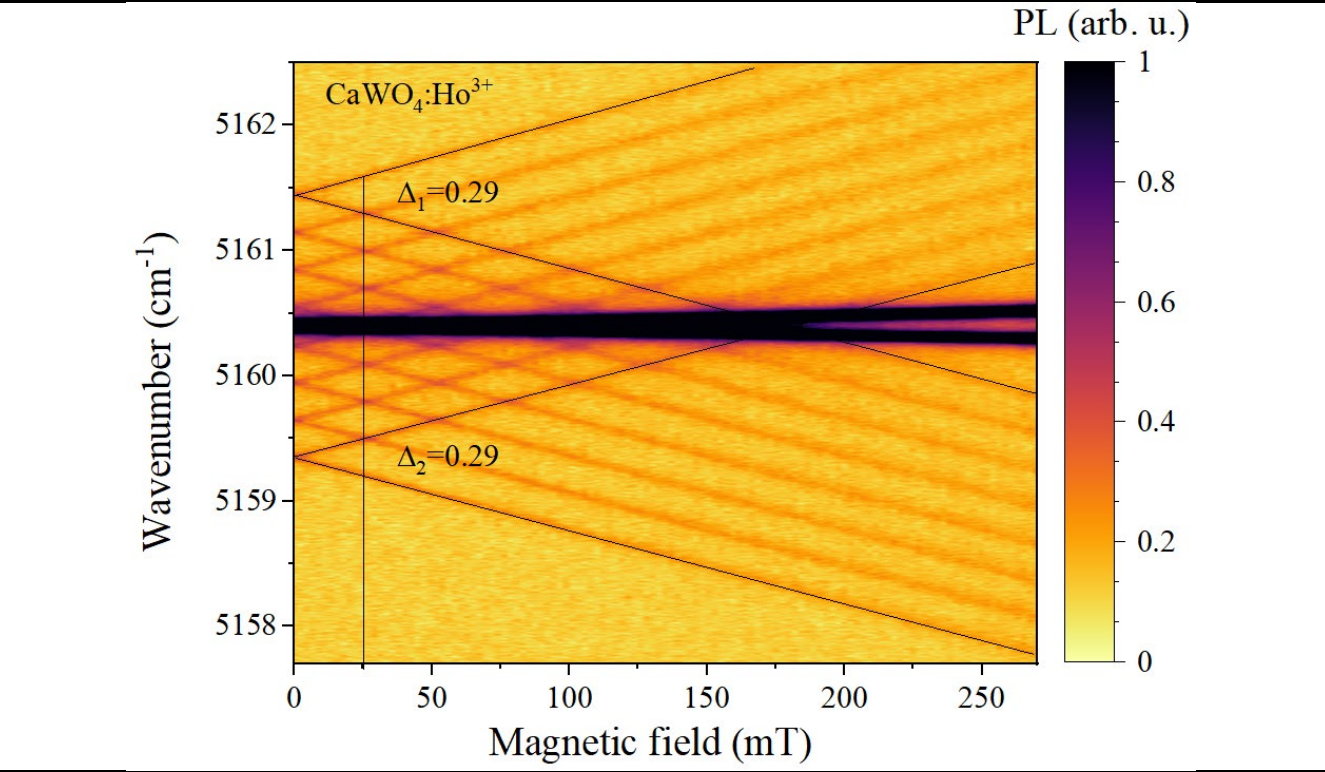

[$^5I_7\,\Gamma_{34}{}^2$ (5160.5) → $^5I_8\,\Gamma_{34}{}^1$ (0)]. $\Delta_{HF}$=2.1 cm$^{-1}$
$g_1 = g_2 = \langle g \rangle = 24.85$

**Fig. S6.** Luminescence intensity maps in the magnetic-field – wavenumber scale for several lines with HFS belonging to the $^5I_7 \rightarrow {}^5I_8$ transition of $Ho^{3+}$ in the PL spectrum of $CaWO_4:Ho^{3+}$ (0.01 at. %) in a magnetic field **B**||***c***. $T$ = 15 K; $\lambda_{ex}$ = 641.6 nm.

## $^5F_5 \rightarrow {}^5I_6$

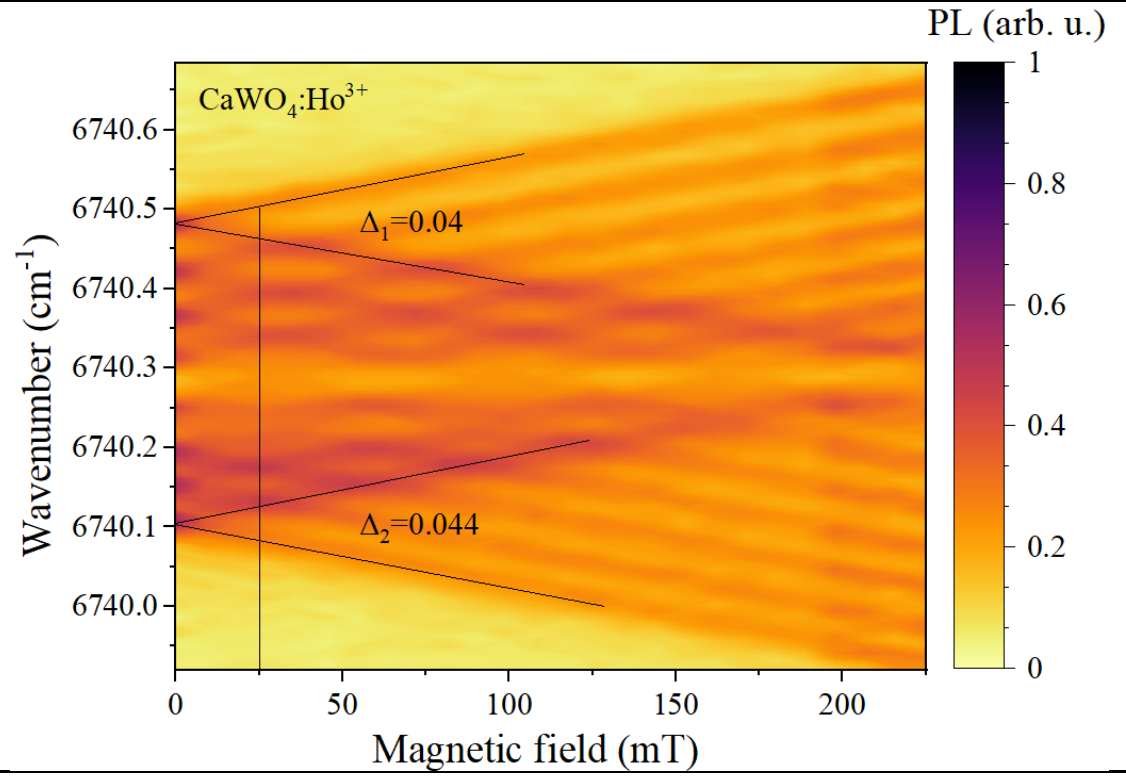

$[{}^5F_5\,\Gamma_2{}^1\,(15410.1) \rightarrow {}^5I_6\,\Gamma_{34}{}^2\,(8669.8)]$.
$\Delta_{HF}$=0.378 cm$^{-1}$
$g_1$ = 3.43, $g_2$ = 3.77; $<g>$ = 3.6

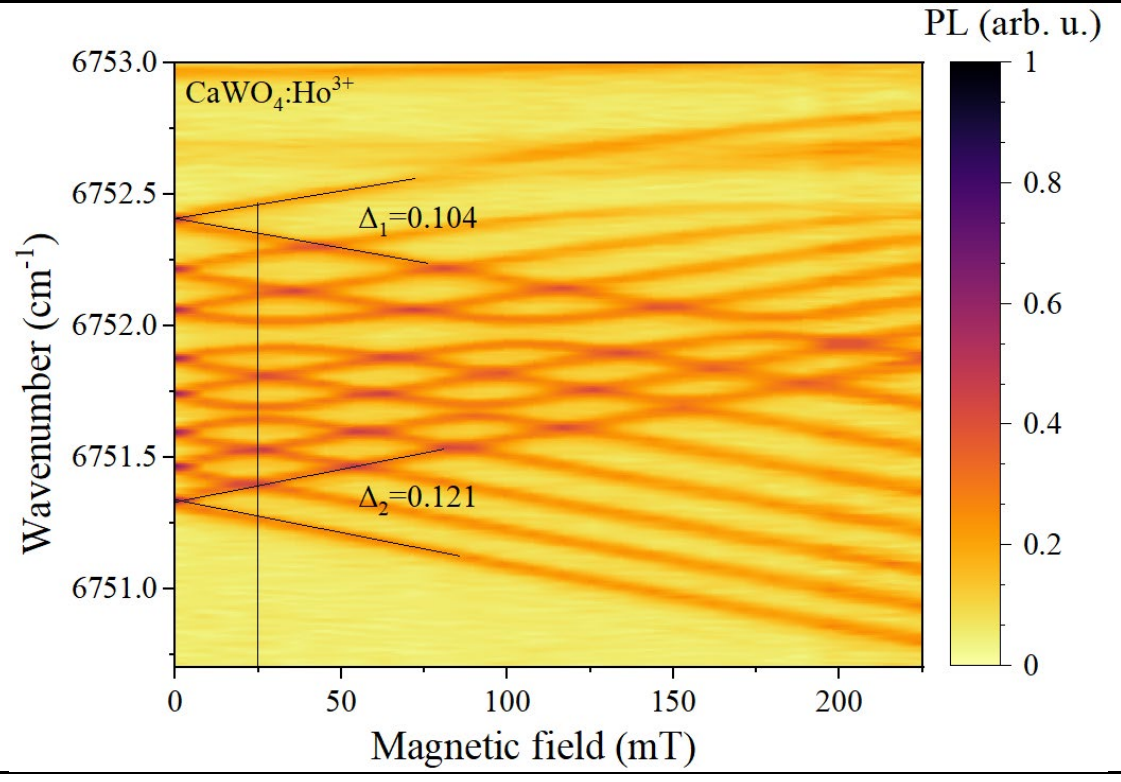

$[{}^5F_5\,\Gamma_2{}^1\,(15410.1) \rightarrow {}^5I_6\,\Gamma_{34}{}^1\,(8658.2)]$.
$\Delta_{HF}$=1.076 cm$^{-1}$
$g_1$ = 8.91, $g_2$ = 10.37; $<g>$ = 9.64

**Fig. S7.** Luminescence intensity maps in the magnetic-field – wavenumber scale for several lines with HFS belonging to the $^5F_5 \rightarrow {}^5I_6$ transition of $Ho^{3+}$ in the PL spectrum of $CaWO_4$:$Ho^{3+}$ (0.01 at. %) in a magnetic field **B**||*c*. $T$ = 15 K; $\lambda_{ex}$ = 641.6 nm.

## $^5I_6 \rightarrow {}^5I_8$ and $^5I_5 \rightarrow {}^5I_7$

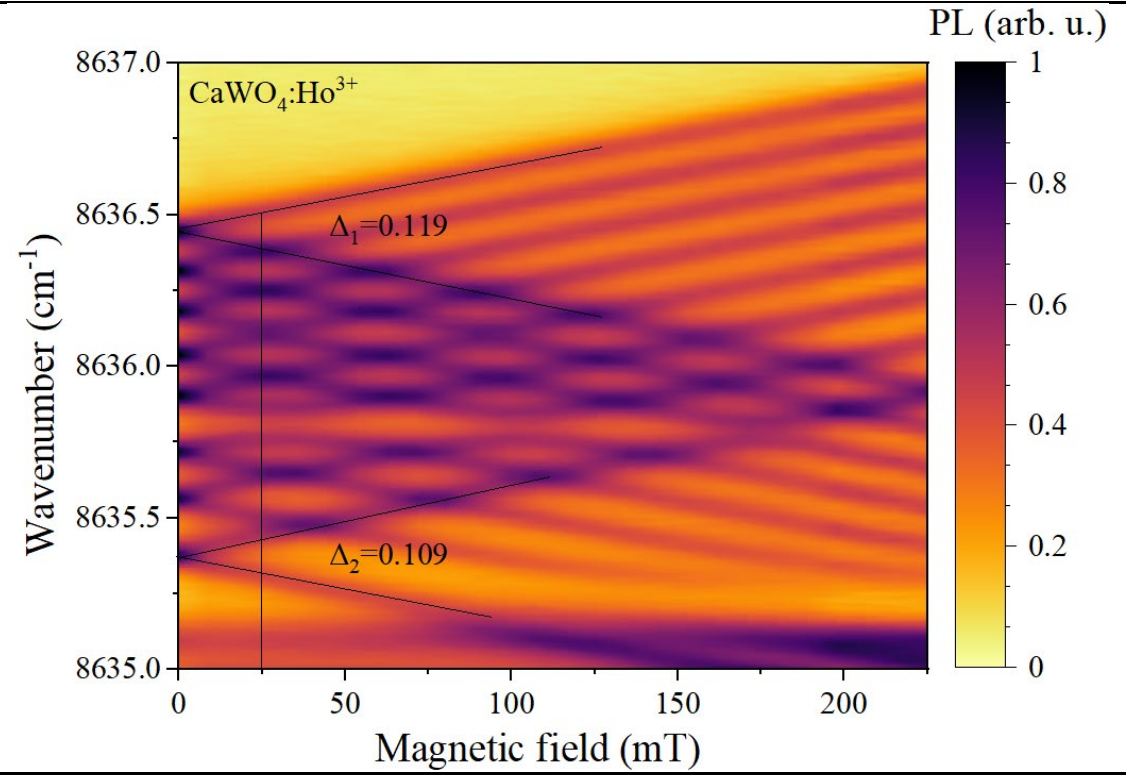

$[{}^5I_6\,\Gamma_{34}{}^1\,(8658.2) \rightarrow {}^5I_8\,\Gamma_2{}^2\,(22.17)]$.
$\Delta_{HF}$=1.082 cm$^{-1}$
$g_1$ = 10.20, $g_2$ = 9.34; $<g>$ = 9.77

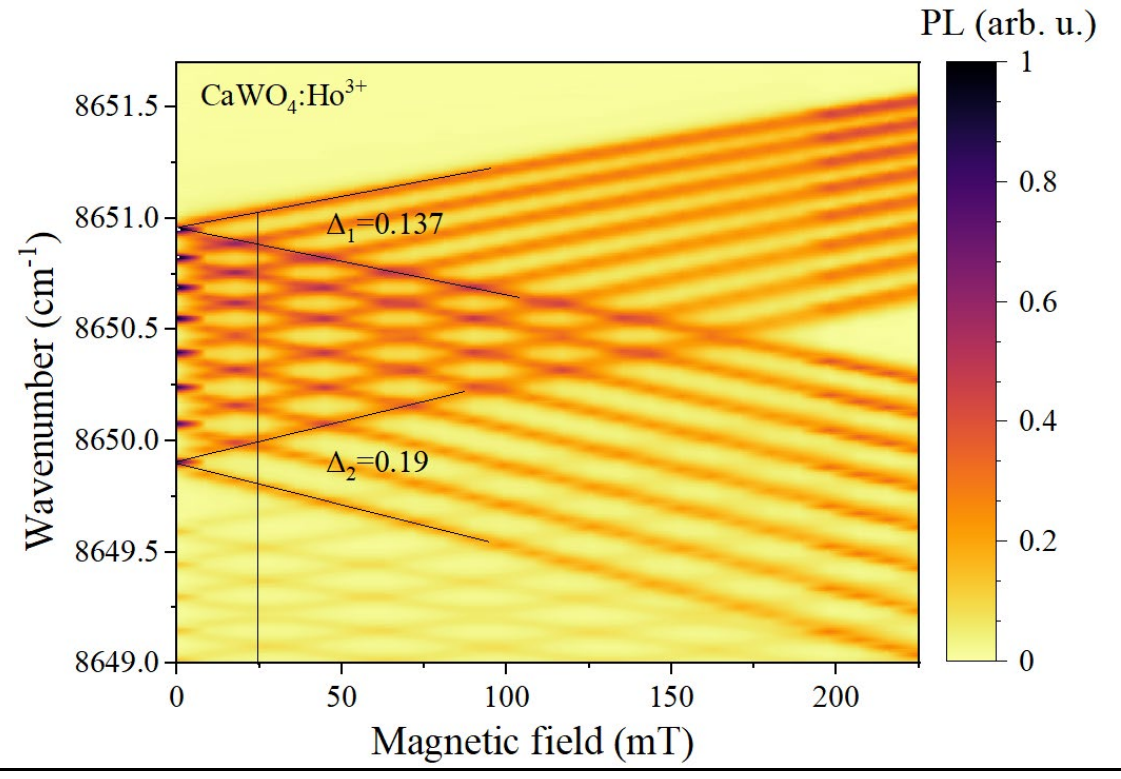

$[{}^5I_6\,\Gamma_2{}^1\,(8650.5) \rightarrow {}^5I_8\,\Gamma_{34}{}^1\,(0)]$.
$\Delta_{HF}$=1.047 cm$^{-1}$
$g_1$ = 11.74, $g_2$ = 16.28; $<g>$ = 14.01

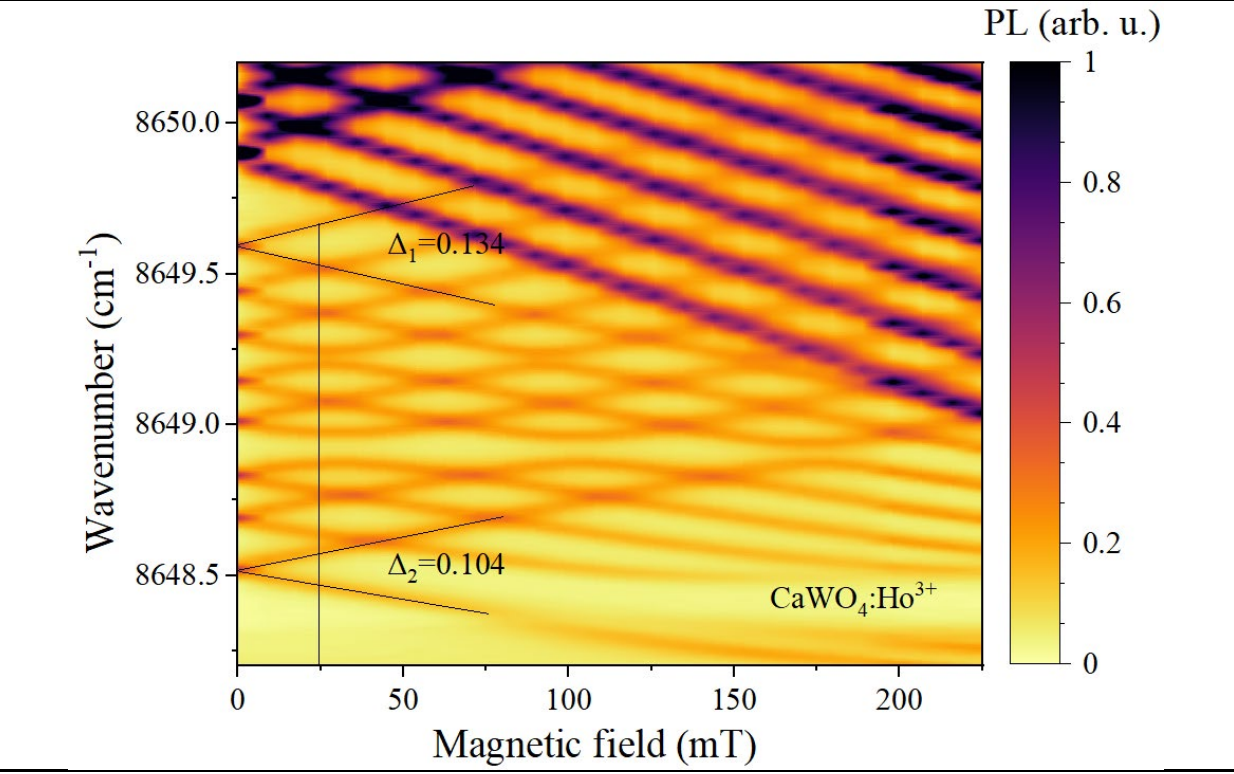

$[^5I_6\,\Gamma_{34}^1\,(8658.2) \rightarrow {}^5I_8\,\Gamma_2^1\,(9.16)]$.
$\Delta_{HF}$=1.074 cm$^{-1}$
$g_1 = 11.48$, $g_2 = 8.91$; $\langle g \rangle = 10.2$

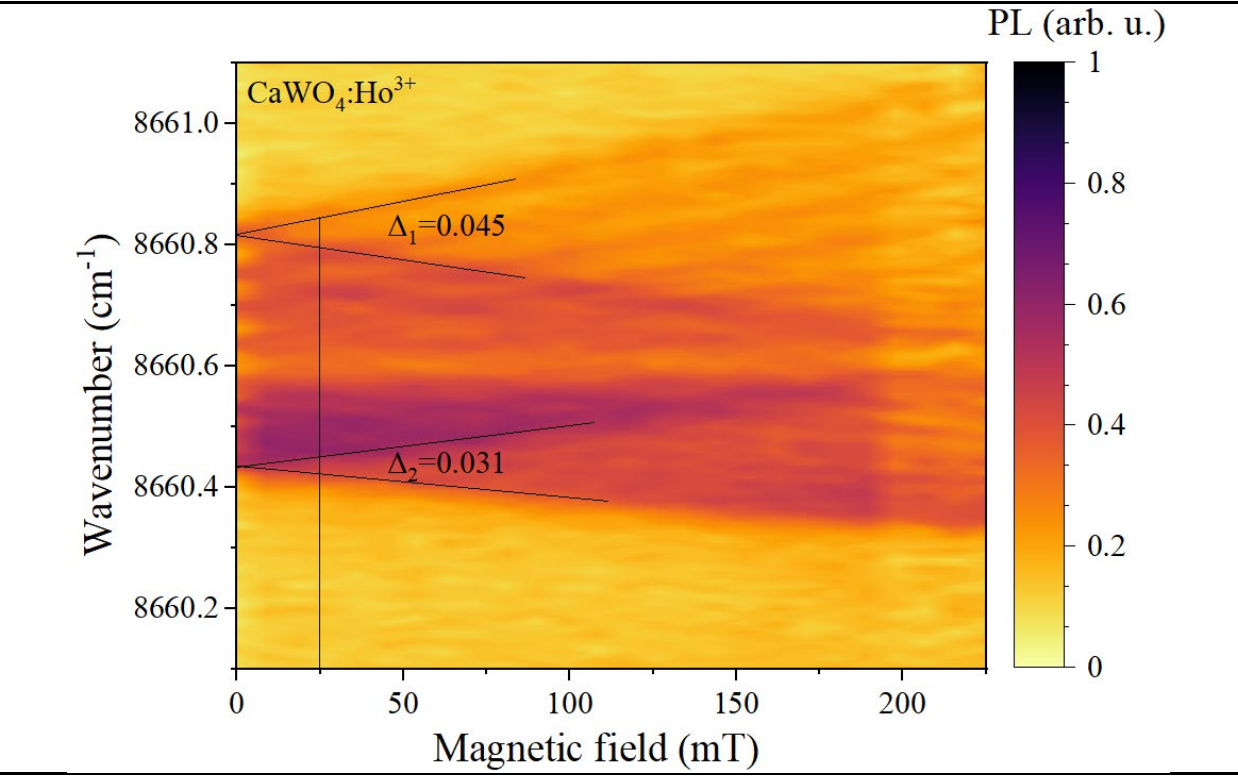

$[^5I_6\,\Gamma_{34}^2\,(8669.8) \rightarrow {}^5I_8\,\Gamma_2^1\,(9.16)]$.
$\Delta_{HF}$=0.386 cm$^{-1}$
$g_1 = 3.86$, $g_2 = 2.66$; $\langle g \rangle = 3.26$

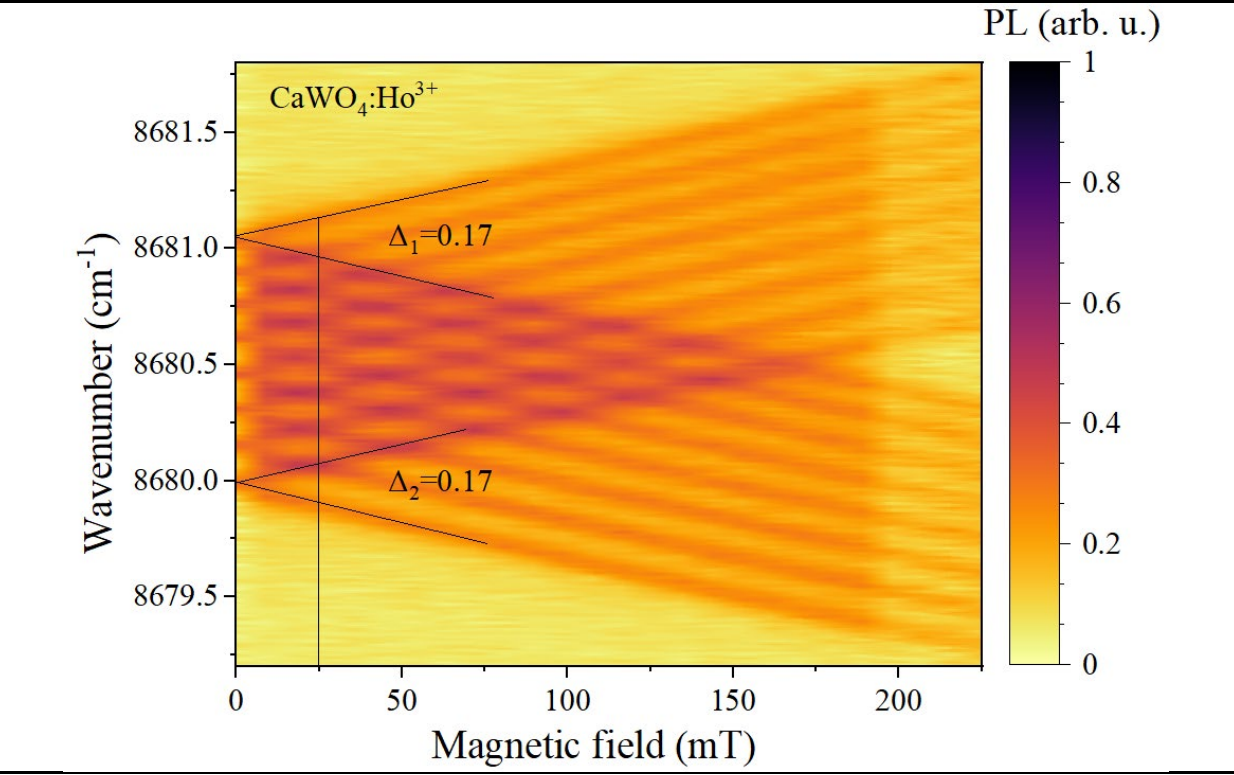

$[^5I_6\,\Gamma_2^3\,(8680.5) \rightarrow {}^5I_8\,\Gamma_{34}^1\,(0)]$.
$\Delta_{HF}$=1.047 cm$^{-1}$
$g_1 = g_2 = \langle g \rangle = 14.57$

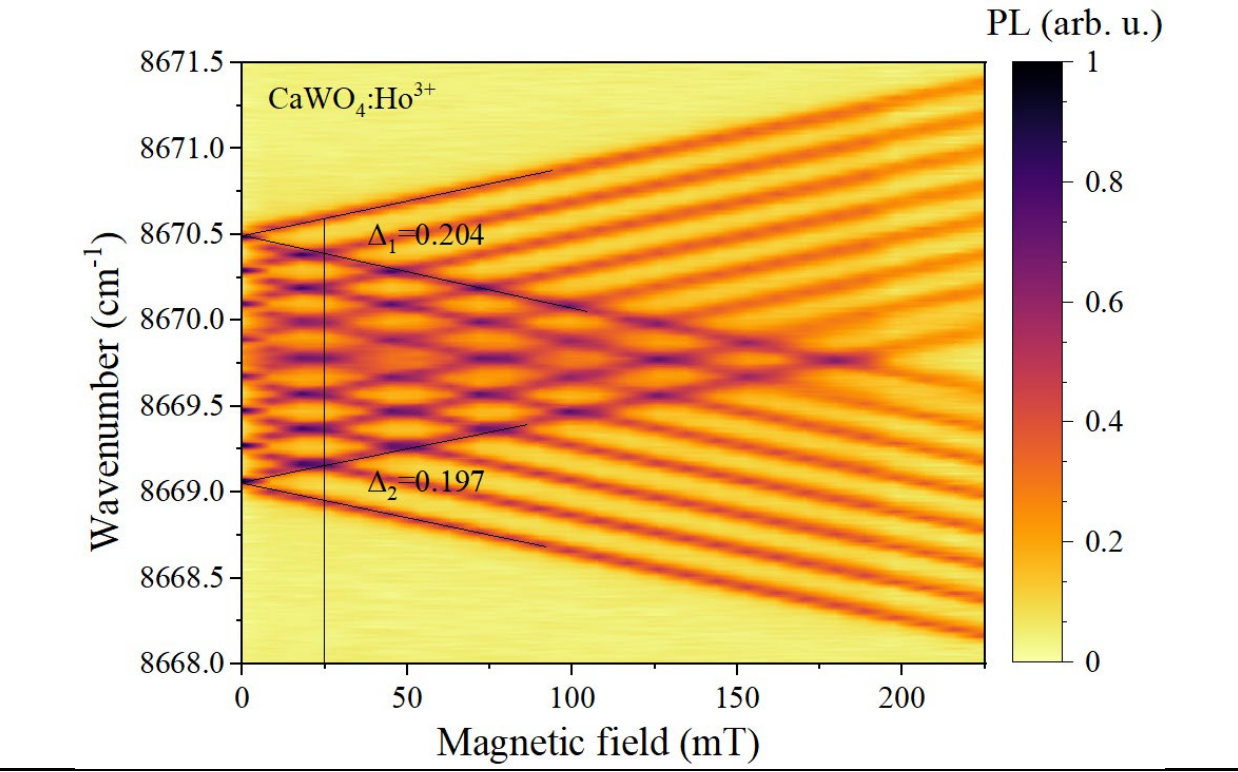

$[^5I_6\,\Gamma_{34}^2\,(8669.8) \rightarrow {}^5I_8\,\Gamma_{34}^1\,(0)]$.
$\Delta_{HF}$=1.43 cm$^{-1}$
$g_1 = 17.48$, $g_2 = 16.88$; $\langle g \rangle = 17.18$

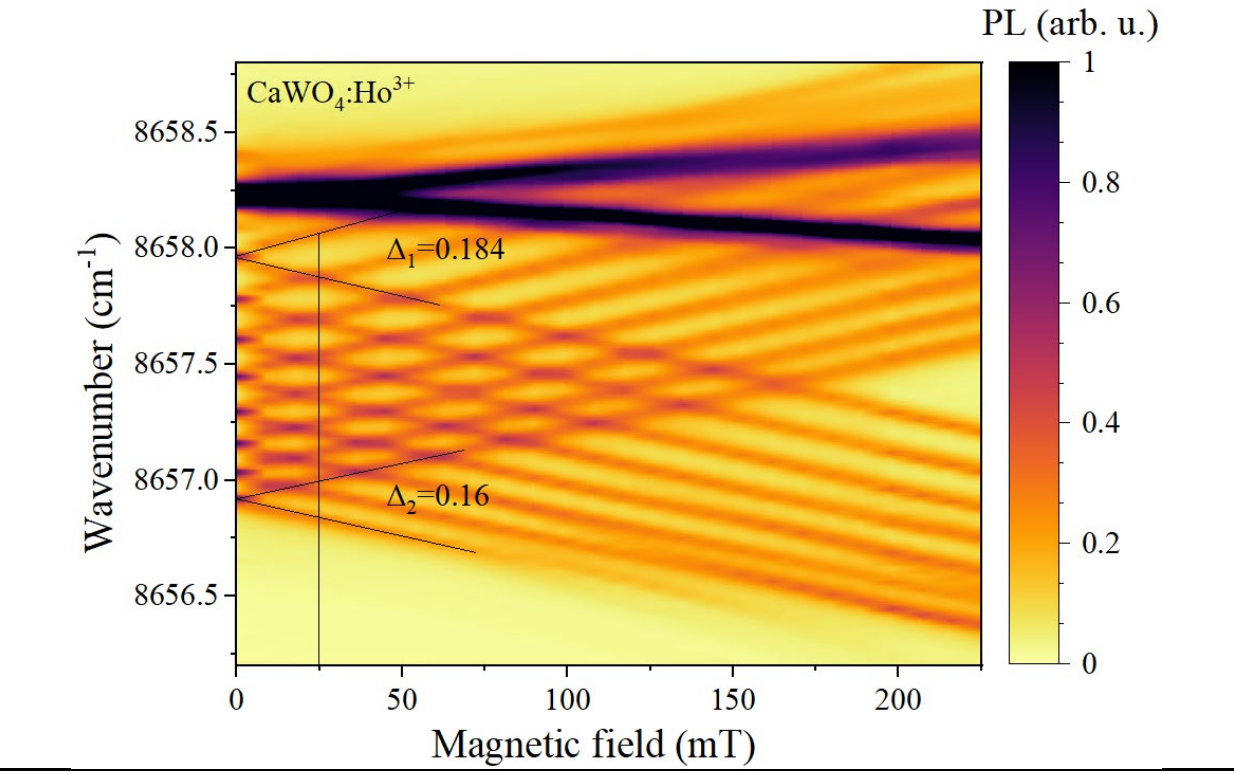

$[^5I_6\,\Gamma_2^2\,(8657.4) \rightarrow {}^5I_8\,\Gamma_{34}^1\,(0)]+$
$[^5I_6\,\Gamma_{34}^1\,(8658.2) \rightarrow {}^5I_8\,\Gamma_{34}^1\,(0)]$.
$\Delta_{HF}$=1.042 cm$^{-1}$
$g_1 = 15.77$, $g_2 = 13.71$; $\langle g \rangle = 14.74$

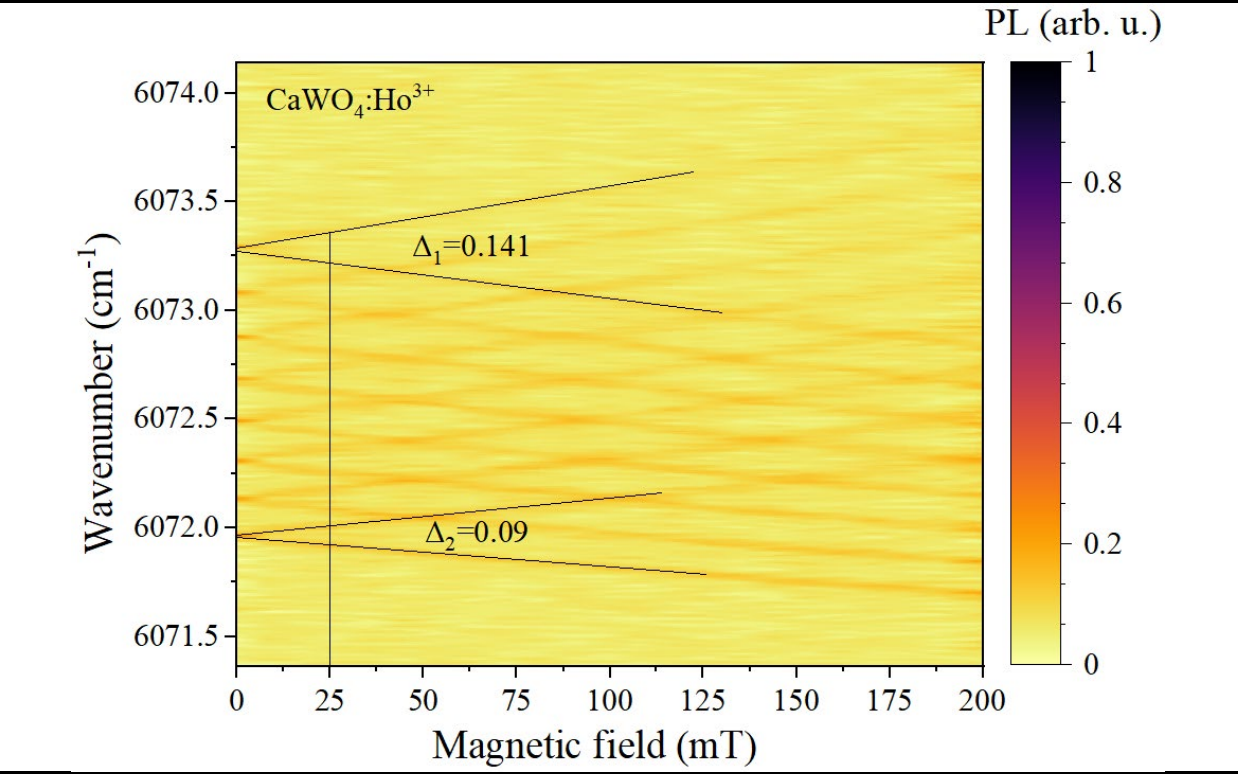

$[^5I_5\,\Gamma_{34}^1\,(11208.9) \rightarrow {}^5I_7\,\Gamma_2^1\,(5137.2)]$
$\Delta_{HF}$=1.319 cm$^{-1}$
$g_1 = 12.08$, $g_2 = 7.71$; $\langle g \rangle = 9.9$

**Fig. S8.** Luminescence intensity maps in the magnetic-field – wavenumber scale for several lines with HFS belonging to the $^5I_6 \rightarrow {}^5I_8$ and $^5I_5 \rightarrow {}^5I_7$ transitions of $Ho^{3+}$ in the PL spectrum of $CaWO_4:Ho^{3+}$ (0.01 at. %) in a magnetic field **B**||***c***. $T$ = 15 K; $\lambda_{ex}$ = 641.6 nm.

## 5. Free-ion parameters for $Ho^{3+}$ in $CaWO_4:Ho^{3+}$

The free-ion Hamiltonian reads:

$$H_0 = \zeta \sum_k \boldsymbol{l}_k \boldsymbol{s}_k + \alpha \hat{\boldsymbol{L}}^2 + \beta \hat{G}(G_2) + \gamma \hat{G}(R_7) + \sum_q (F^q \hat{f}_q + P^q \hat{p}_q + T^q \hat{t}_q + M^q \hat{m}_q), \qquad \text{(S1)}$$

where $\zeta$ is the single-electron spin-orbit interaction constant; $F^{(2)}$, $F^{(4)}$, and $F^{(6)}$ are Slater integrals which determine the electrostatic interactions between 4*f* electrons and lead to the formation of electronic multiplets $^{2S+1}L_J$; additional parameters α, β, γ, $T^{(q)}$ and $P^{(q)}$ ($q$=2,4,6), $M^{(q)}$ ($q$=0,2,4) are introduced to account for the shifts of 4*f*-multiplets due to electrostatic interactions between the ground and excited electronic configurations and electrostatically correlated spin-orbit and magnetic interactions, respectively. The following values (in $cm^{-1}$) were used: $\zeta$=2133.3, $F^{(2)}$=93668, $F^{(4)}$=66113, $F^{(6)}$=49372, α = 18.9, β = - 611, γ = 2013, $P^{(2)}$ = 528, $P^{(4)}$ = 396, $P^{(6)}$ = 264, $T^{(2)}$ = 249, $T^{(3)}$ = 37, $T^{(4)}$ = 98, $T^{(6)}$ = - 316, $T^{(7)}$ = 440, $T^{(8)}$ = 372, $M^{(0)}$ = 3, $M^{(2)}$ = 1.7, and $M^{(4)}$ = 1.1.